\documentclass[lettersize,journal]{IEEEtran}
\usepackage{amsmath,amssymb,amsfonts}
\usepackage{algorithmic}
\usepackage{algorithm}
\usepackage{array}
\usepackage{caption}
\usepackage[caption=false,font=normalsize,labelfont=sf,textfont=sf]{subfig}
\usepackage{textcomp}
\usepackage{stfloats}
\usepackage{url}
\usepackage{verbatim}
\usepackage{graphicx}
\usepackage{cite}
\usepackage{booktabs}
\usepackage[compact]{titlesec}
\titlespacing{\section}{0pt}{*0.9}{*0.5}
\titlespacing{\subsection}{0pt}{*0.7}{*0.4}
\titlespacing{\subsubsection}{0pt}{*0.6}{*0.3}

\begin{document}

\newcommand{\SYSNAME}{Arca}

\title{\SYSNAME: A Lightweight Confidential Container Architecture for Cloud-Native Environments}

\author{Di~Lu,~\IEEEmembership{Member,~IEEE,}
Mengna~Sun,
Qingwen~Zhang,
Yujia~Liu,
Jia~Zhang,
Xuewen~Dong,~\IEEEmembership{Member,~IEEE,}
Yulong~Shen,~\IEEEmembership{Member,~IEEE,}
and~Jianfeng~Ma,~\IEEEmembership{Member,~IEEE}
\thanks{\textbullet\ Di Lu, Mengna Sun, Qingwen Zhang, Yujia Liu, Xuewen Dong, and Yulong Shen are with the School of Computer Science and Technology, Xidian University, Xi’an, Shaanxi 710071, China, and also with the Shaanxi Key Laboratory of Network and System Security, Xi'an, Shaanxi 710071, China. 
E-mail: \{dlu, xwdong\}@xidian.edu.cn, smn20200419@163.com, 15319018420@163.com, 19829259628@163.com, ylshen@mail.xidian.edu.cn.\\\textbullet\ Jia Zhang is with Alibaba Cloud, Hangzhou, China. E-mail: qianyue.zj@alibaba-inc.com.\\
\textbullet\ Jianfeng Ma is with the School of Cyber Engineering, Shaanxi Key Lab of Network and System Security, Xidian University, Xi’an, China. E-mail: jfma@mail.xidian.edu.cn.}%
\thanks{Manuscript received April 19, 2021; revised August 16, 2021.}}

\markboth{Journal of \LaTeX\ Class Files,~Vol.~14, No.~8, August~2021}%
{Shell \MakeLowercase{\textit{et al.}}: A Sample Article Using IEEEtran.cls for IEEE Journals}

\maketitle

\begin{abstract}
Confidential containers protect cloud-native workloads using trusted execution environments (TEEs). However, existing Container-in-TEE designs (e.g., Confidential Containers(CoCo)) encapsulate the entire runtime within the TEE, inflating the trusted computing base (TCB) and introducing redundant components and cross-layer overhead.
We present \SYSNAME, a lightweight confidential container framework based on a TEE-in-Container architecture that isolates each workload in an independent, hardware-enforced trust domain while keeping orchestration logic outside the TEE.
This design minimizes inter-layer dependencies, confines compromise to per-container boundaries, and restores the TEE’s minimal trust principle.
We implemented \SYSNAME~on Intel SGX, Intel TDX, and AMD SEV.
Experimental results show that \SYSNAME~achieves near-native performance and outperforms CoCo in most benchmarks, while the reduced TCB significantly improves verifiability and resilience against host-level compromise.
\SYSNAME~demonstrates that efficient container management and strong runtime confidentiality can be achieved without sacrificing security assurance.
\end{abstract}


\section{Introduction}\label{sec:intro}
\IEEEPARstart{D}{ifferent}  from the traditional cloud computing model, cloud-native architecture\cite{CNCF}-\cite{CloudNative} is an approach that designs and builds applications to take advantage of the elasticity and distributed nature of the cloud. 
Typically, cloud-native applications are broken down into multiple, self-contained services using technologies and methodologies: microservices, containers, continuous integration \& continuous delivery (CI/CD), DevOps, and declarative APIs, enabling teams to deploy and scale components independently to make updates, fix issues, and deliver new features without any service interruption.
This means cloud-native applications make the most of modern infrastructure's dynamic, distributed nature to achieve greater speed, agility, scalability, reliability, and cost efficiency. 

As one of the key pillars of cloud-native architecture, containers\cite{DockerDocs} are lightweight, executable components with all the elements needed—including application source code and dependencies—to run the code in any environment. Containers provide workload portability, enabling ``build once, run anywhere'' code, making development and deployment significantly easier. They also help reduce friction between languages, libraries, and frameworks since they can be deployed independently. This portability and flexibility make containers ideal for building microservice architectures.  

Unlike virtual machines (VMs), containers run directly on the host system (sharing its kernel) and do not need to emulate devices or maintain large disk files. 
Further, according to the OCI (Open Container Initiative) Distribution Specification, the dependencies used to spawn or deploy containers are specified as layers that can be shared between different ones. This makes them amenable to caching, speeding up deployment while reducing storage costs for multiple containers. 
However, the lower isolation and the flexible resource sharing pose significant runtime security challenges for containerized applications \cite{Dockersecure1}-\cite{Dockersecure2}.
For example, attackers can compromise a host running containers, steal sensitive data, and tamper with code to compromise the confidentiality and integrity of critical applications. In addition, the dependencies (such as image files) shared among containers may contain malicious code, increasing the risk of in-container applications and data privacy leakage. 

To address the security issues above, TEE technology\cite{Sabt2015}-\cite{McKeen2013} is introduced to provide a strongly isolated runtime environment for running critical code and protecting sensitive data processing, even if the host system is compromised or controlled by malicious entities. TEEs are available in mainstream CPU vendors, either process-based, such as Intel SGX\cite{Costan2016}, ARM TrustZone\cite{Pinto2019}, and RISC-V Keystone\cite{Lee2020}, or VM-based, such as AMD SEV-SNP\cite{Kaplan2016}-\cite{Sev-Snp2020}, Intel TDX\cite{Cheng2024}, and ARM CCA\cite{Li2022}, which offer hardware-level isolation of the VM, preventing the host operating system and the hypervisor from accessing the VM’s memory and registers.

By adopting TEE technologies, CoCo\cite{CoCo}, a well-developed open-source containerized confidential computing project, is proposed to provide a built-in TEE container environment for running critical code and processing sensitive data. This becomes an ideal solution that ensures the confidentiality and integrity of containerized applications in a cloud-native environment. 
However, due to migrating containers and the relevant dependencies into TEE, confidential containers like CoCo have several non-negligible shortcomings that significantly affect the security, efficiency, and availability of protected containerized applications:
\begin{enumerate}
    \item Potential security issue: supporting containerized workloads in TEE introduces additional components, services, and code, inflating the TCB. However, with the TEE's functions becoming richer, its attack surface has also been enlarged, making it more vulnerable and difficult to secure, audit, and measure\cite{McCune2008, Naila2019, Lily2021}.

    \item Extra performance cost: An inflated TCB with rich components and functions can cost more computing resources due to running and retaining too much extra code that is not directly relevant to the security functionalities of the TCB\cite{McCune2008}.
\end{enumerate}

To address the issues above, we propose \SYSNAME\footnote{The source code of \SYSNAME, including implementation and scripts, is available at our GitHub repository: https://github.com/nijino2002/Arca.}. This new confidential container architecture focuses on safeguarding critical processes (running code) with a TEE rather than migrating a runtime environment, such as a container and its dependencies, into the TEE, thereby making the TCB significantly lighter than CoCo-like systems. 
The contributions of our work are listed as follows:
\begin{enumerate}
    \item A brand-new, generic, confidential computing architecture, \SYSNAME, is proposed for the cloud-native scenario, implementing an in-container TEE environment that achieves higher efficiency and effectively prevents the TCB from being inflated compared to conventional approaches.

    \item We implemented \SYSNAME{} using three mainstream cloud-native TEE technologies---Intel SGX, Intel TDX, and AMD SEV —in the Alibaba cloud-native environment to demonstrate the feasibility and availability of the proposed architecture.

    \item We evaluated \SYSNAME, implemented with different TEE technologies, and compared its performance with CoCo. Finally, the security discussion on \SYSNAME{} is presented.
\end{enumerate}

This paper is organized as follows.
A brief introduction to cloud-native and its confidential computing technologies is given in Section~\ref{sec:bg}. Section~\ref{sec:threat_assumption} clarifies the threat model and security assumptions. After that, we detail the design of \SYSNAME~in Section~\ref{sec:sys_dsgn}. Based on our design, the essentials of implementing the key components and functionalities in \SYSNAME~are detailed in Section~\ref{sec:sys_imp}. Section~\ref{sec:perf_eval} first gives the experimental setup. Then, the system performance evaluations are presented and discussed, followed by a security analysis of \SYSNAME in Section \ref{sec:sec_dis}. Section~\ref{sec:rel_work} discusses the typical research work on confidential containers under the cloud-native environment. Finally, we conclude the paper in Section~\ref{sec:concl}.

\section{Background}\label{sec:bg}

\subsection{Cloud Native}\label{subsec:bg-cn}

Cloud-native architecture is a design and development approach that fully leverages the elasticity, scalability, and high availability of cloud computing platforms. Unlike traditional cloud computing architectures, which primarily rely on virtual machines or monolithic structures, cloud-native architectures leverage technologies such as microservices, containerization, continuous integration and continuous delivery (CI/CD), DevOps, and declarative APIs. These technologies decompose applications into multiple loosely coupled service units, enhancing flexibility and maintainability. Moreover, cloud-native architecture emphasizes automated deployment, elastic scaling, and self-healing capabilities, enabling applications to adapt to changing market demands quickly and to efficiently migrate across multi-cloud or hybrid cloud environments, thereby reducing dependence on specific cloud service providers.  

Despite the advantages of agility and scalability offered by cloud-native architecture, security remains a critical challenge\cite{Cloudsecurity}-\cite{Cloud_security}. Cloud-native environments extensively adopt containerization and microservice architectures, with these components typically running in dynamic, distributed environments, significantly expanding the potential attack surface. In particular, containerized applications face a range of security risks due to their lower isolation and flexible resource-sharing characteristics. In multi-tenant cloud environments, containers share the host operating system kernel, which can result in insufficient isolation between containers, thereby elevating the risk of security vulnerabilities\cite{Kernel_security}. Consequently, sensitive data and executable code running within containers may become susceptible to unauthorized access by the host system or other containers. Additionally, the shared dependencies between containers may harbor malicious code, further amplifying the threat of data breaches and compromising privacy\cite{Image_security}. Therefore, securing containerized applications, especially in multi-tenant cloud environments, has emerged as a critical challenge.  

\subsection{Confidential Computing for Cloud Native}\label{subsec:bg-cccn}

Confidential computing technology ensures the confidentiality and integrity of sensitive data and code during computation by establishing a hardware-level, isolated execution environment. This environment safeguards data and codes against unauthorized access or tampering, even in untrusted settings. Specifically, TEE utilizes hardware-based isolation techniques, such as Intel TDX, Intel SGX, and AMD SEV, to create a secure enclave where sensitive data remains encrypted during execution, preventing unauthorized actors or malicious software from accessing or altering it. 

In cloud-native architectures, TEE leverages its hardware isolation to establish a secure computational domain for applications within a cloud environment. This ensures sensitive data is protected from both malicious attacks and unauthorized access during processing. Cloud-native platforms typically operate within multi-tenant environments, where applications and services share underlying hardware resources. The hardware-level isolation provided by TEEs effectively mitigates security vulnerabilities between containers, ensuring that even if the host machine is compromised, applications running in containers remain securely isolated within the protected environment. Consequently, integrating confidential computing technology with containerization to create a "confidential container" offers an innovative solution for secure computation in containerized environments. Confidential containers significantly enhance the runtime security of both code and data within containers, while also providing a more robust, resilient TCB for cloud-native applications. 

However, existing solutions for confidential containers\cite{Brasser2022}-\cite{Lee2024} face several challenges, including TCB expansion, an increased attack surface, and non-negligible performance overhead. Moreover, their development, deployment, and maintenance processes are often complex and impose substantial technical barriers, significantly hindering the widespread adoption of confidential containers and the growth of their ecosystem. 

To better understand the performance implications of different TEE substrates in cloud-native environments, we next compare \emph{process-based} and \emph{VM-based} TEEs, which represent two mainstream architectures underlying confidential container systems.

\subsection{Process-based and VM-based TEEs}
\begin{figure}
    \centering
    \includegraphics[width=0.85\linewidth]{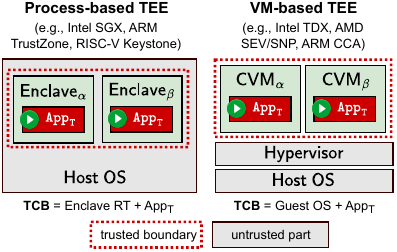}
    \caption{Conceptual distinction between process-based and VM-based TEEs,
highlighting their differences in protection scope, trust boundary, and TCB size.}
    \label{fig:tee_types}
\end{figure}

Currently, TEE can be broadly categorized into two architectural families: \emph{process-based} and \emph{VM-based}. 
As illustrated in Fig.~\ref{fig:tee_types}, these two paradigms differ fundamentally in their isolation granularity, trust boundaries, and runtime interaction models.

Process-based TEEs, represented by Intel~SGX, ARM~TrustZone, and RISC-V~Keystone, create secure enclaves within the user space of an untrusted operating system. 
The enclave runtime and the application code form a small, auditable TCB, while the host OS and kernel services remain outside the trusted boundary. 
Although this fine-grained isolation provides strong guarantees for data confidentiality and integrity, it requires frequent enclave transitions (\texttt{ECALL}/\texttt{OCALL}) for system calls and I/O operations. 
These transitions introduce context-switch overhead and complicate I/O handling, which collectively impact performance and scalability.

In contrast, VM-based TEEs—such as Intel~TDX, AMD~SEV/SNP, and ARM~CCA—protect an entire guest virtual machine, including its operating system and applications, as a unified trusted domain under an untrusted hypervisor. 
This coarse-grained isolation extends the TCB to encompass the guest OS, eliminating the need for frequent enclave transitions and reducing host-kernel dependency. 
Hardware-level memory encryption and VM-entry protection ensure that even a compromised host or hypervisor cannot access or tamper with the VM’s internal state.

Overall, the two paradigms reflect a trade-off between \emph{granularity} and \emph{scope}: process-based TEEs favor minimal TCB and strong compartmentalization at the cost of transition latency, whereas VM-based TEEs provide broader isolation with lower runtime overhead but a larger trusted base. 
This structural contrast underpins the performance and security boundary discussions presented in Sections~\ref{sec:perf_eval} and~\ref{sec:sec_dis}.

\section{Threat Model and Assumptions} \label{sec:threat_assumption}
\subsection{Threat Model}\label{subsec:threat_model}
We assume a powerful remote attacker without physical access to the host platform. The attacker is capable of exploiting vulnerabilities in the host operating system or container runtime to escalate privileges to host-level root access. Once the host OS is compromised, the attacker can tamper with container managers such as Docker, manipulate system services, and arbitrarily inspect or modify the memory state of any containerized workload. This enables both direct and indirect compromise of sensitive application assets.

The primary security risks in a container-only environment are summarized below:
\begin{enumerate}
    \item \emph{Threats to Confidentiality}: 
    The attacker may extract sensitive code and data from the memory of containerized applications by bypassing OS-level isolation via host privilege escalation. These actions violate data confidentiality in multi-tenant environments and may lead to the leakage of secrets across cloud tenants.

    \item \emph{Threats to Integrity}: 
    With full control of the host OS, the attacker can arbitrarily modify application binaries, runtime states, or intermediate computation results within a container. This may corrupt the integrity of critical services and influence the correctness of security-sensitive computation.
\end{enumerate}

This threat model aligns with real-world cloud adversaries where operating-system compromise is a plausible assumption, and thus requires a hardware-rooted execution protection scheme for sensitive workloads.

\subsection{Assumptions}\label{subsec:threat_assum}

We assume the attacker’s main objective is to compromise tenant workloads, rather than physically damage the platform. We trust the CPU-provided hardware security primitives and the TEE implementation, which form the root of trust of the entire system, including memory-encryption engines and enclave/VM isolation hardware. We also assume the TEE firmware and trusted computing base (TCB) are properly provisioned and free from vulnerabilities that the adversary can exploit.

Containers are assumed to behave honestly and do not voluntarily disclose their protected assets. However, attempts from compromised co-resident containers to steal or tamper with sensitive data are considered in scope. Side-channel attacks (e.g., cache timing, speculative execution, microarchitectural leaks) and denial-of-service attacks are considered out of scope, as they require either advanced physical capabilities or platform-specific countermeasures. These threats can be mitigated by adopting more resistant TEE architectures if required.

\section{System Design}\label{sec:sys_dsgn}
\subsection{System Architecture}\label{subsec:sys_arch}
\begin{figure}
    \centering
    \includegraphics[width=.7\linewidth]{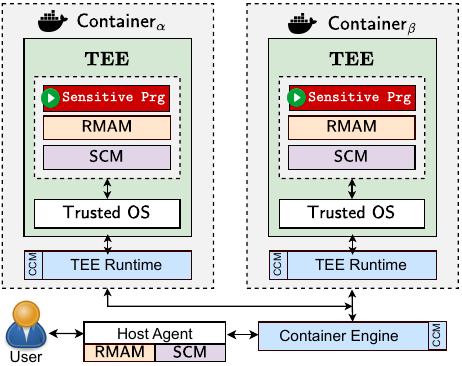}
    \caption{Overall architecture of \SYSNAME{}, which implements a TEE-in-Container model.
Each container embeds an independent hardware-protected TEE instance for secure computation.}
    \label{fig:sys_arch}
\end{figure}

As outlined in Section \ref{sec:intro}, this paper proposes an in-container TEE model, referred to as ``\emph{One Container vs. One TEE (1Cv1T)}'', to achieve high performance and streamlined management. This model enables lightweight and efficient TEE deployment and management by abstracting complex TEE-oriented operations into simplified container operations. Figure \ref{fig:sys_arch} illustrates the architecture of \SYSNAME{} and its workflow, which consists of three core components: the \emph{Confidential Container Manager (CCM)}, the \emph{Secure Communication Module (SCM)}, and the \emph{Remote Attestation Module (RMAM)}. We will briefly introduce these modules by describing a typical workflow in the figure.

Firstly, when a user submits a confidential task to the platform, the CCM orchestrates the deployment of a container. To ensure the trustworthiness of the TEE and its platform, the RMAM performs \emph{trust measurement}—verifying the authenticity and integrity of the TEE provider---both before the TEE is instantiated and continuously during its execution. Once the TEE platform is successfully attested, the CCM initializes an in-container TEE environment, where the sensitive program and data are securely extracted and decrypted for confidential execution. Following the establishment of the in-container TEE, the SCM facilitates secure communication between the user application in the untrusted environment and the confidential workload within the TEE, ensuring data integrity and confidentiality throughout the execution lifecycle.

The subsequent sections provide a detailed discussion of each module's functionality.
\subsection{Confidential Container Manager (CCM)}
As previously described, the CCM is responsible for the construction and management of both the confidential container and the TEE executing within it. Upon receiving a sensitive task submission, the CCM first instantiates a container and provisions an in-container TEE environment, such as a confidential virtual machine (CVM) based on AMD SEV. Subsequently, the confidential task's code and data are securely extracted and decrypted within the TEE to ensure secure execution. For instance, when AMD SEV is designated as the cloud-native TEE provider, the task is deployed and executed within an SEV-enabled CVM, preserving the integrity and confidentiality of the computation. Since the CCM establishes a TEE space within a container, the TEE can focus solely on executing confidential tasks. This approach contrasts with the traditional confidential container scheme, which embeds a container inside the TEE, requiring additional runtime libraries that significantly increase the TEE's size and compromise the minimal TCB principle.

\subsection{Secure Communication Module (SCM)}\label{subsec:sec_comm}
Upon initializing the TEE, \SYSNAME~executes the user-specified sensitive program within the secure enclave. At this stage, the program's input and output remain confined within the TEE, preventing direct user interaction. To overcome this limitation, \SYSNAME{} employs a secure communication module that enables the redirection of input and output between the TEE and the host machine. This mechanism ensures seamless, confidential interaction between user applications running on an untrusted host and the sensitive program executing within the TEE, thereby preserving data integrity and confidentiality throughout the transmission process.

\subsection{Remote Attestation Module (RMAM)}\label{subsec:remote_att}
The RMAM verifies the authenticity of the platform on which the TEE depends and the integrity of its execution environment, ensuring that sensitive workloads run exclusively in trusted environments. To achieve this, RMAM leverages the native TEE’s remote attestation mechanism to perform comprehensive trust measurements and report critical system states. By exporting the platform’s certificate chain and generating encrypted integrity reports for the TEE, RMAM enables remote parties to securely validate the platform's authenticity and the integrity of the execution environment before and during workload execution, mitigating the risks of unauthorized modification and adversarial interference. 

\subsection{Key Management}\label{subsec:key_mgmt}
Key management in \SYSNAME{} follows a general principle of derivation rather than storage, which aligns with the design philosophy of mainstream TEEs. Instead of persistently storing cryptographic material, \SYSNAME~derives all operational keys dynamically from hardware-protected roots and runtime measurements, ensuring that cryptographic states remain tightly coupled with the platform’s integrity and the confidentiality of its execution environment.

At the hardware layer, each TEE-capable processor embeds a non-extractable \emph{Root of Trust for Key Derivation (RoT-KD)}—a hardware-fused secret provisioned at manufacturing time. This root secret never leaves the processor boundary and serves as the seed for an internal key derivation function (KDF). When a confidential container is initialized, the KDF combines this root secret with context-specific attributes—such as the TEE’s measurement hash, security version, and the container identity—to derive a hierarchy of working keys. These include the \emph{attestation key} for platform identity verification, the \emph{sealing key} for data persistence protection, and the \emph{session key} for runtime communication confidentiality.

The lifecycle of keys in \SYSNAME~is entirely ephemeral. Derived keys exist only in volatile memory within the TEE boundary and are destroyed when the container or the TEE instance terminates. When persistent data protection is required, \SYSNAME{} employs a sealing mechanism: rather than storing the key itself, it encrypts data objects with a measurement-bound sealing key and stores the ciphertext externally. Upon reinitialization under the same measurement context, the same key can be deterministically re-derived to decrypt the sealed data.

This derivation-centric design yields two security advantages. First, the absence of long-lived plaintext keys eliminates the attack surface associated with key storage and exfiltration. Second, binding all keys to verifiable system measurements automatically revokes them upon any modification to the software stack or TEE configuration, ensuring that cryptographic trust is inseparable from the attested system state. Consequently, the key management framework in \SYSNAME{} provides a unified and context-dependent trust foundation that scales across heterogeneous TEE architectures and containerized workloads.

To illustrate this process concretely, consider a deployment of \SYSNAME~on an Intel TDX-enabled confidential virtual machine. During the creation of a Trust Domain (TD), the TDX module derives several domain-specific keys from a chip-fused root secret using an internal hardware KDF. Among these, a \emph{TD-SEAL key} is derived by combining the root secret with the TD’s measurement hash (\texttt{MRTD}) and configuration attributes. When the confidential container needs to persist sensitive data, it uses this TD-SEAL key to encrypt the data into a sealed blob stored outside the TEE. Because the TD-SEAL key can be deterministically re-derived only when the TD is recreated with an identical measurement, any modification to the container image or TEE configuration renders the sealed data undecryptable. Similarly, when establishing a secure channel to the remote attestation service, the TD generates an ephemeral session key through an ECDH exchange bound to its attested public key; the session key resides solely inside the TD and is destroyed once the attestation completes. This concrete example shows how the abstract derivation-based model in \SYSNAME{} naturally maps to practical hardware mechanisms, ensuring that every key used in the system is both state-dependent and transient. 

\section{TEE-Dependent \SYSNAME{} Implementation} \label{sec:sys_imp}

To demonstrate the feasibility and practicality of the \SYSNAME{} architecture, we have implemented it across three mainstream cloud-native TEE technologies: Intel SGX, Intel TDX, and AMD SEV. In the following sections, we provide a detailed discussion of the platform-specific design considerations and the key aspects in implementing the core components---CCM, SCM, and RMAM.

\subsection{\SYSNAME~for Intel SGX}\label{subsec:imp_sgx}
\begin{figure}
    \centering
    \includegraphics[width=0.75\linewidth]{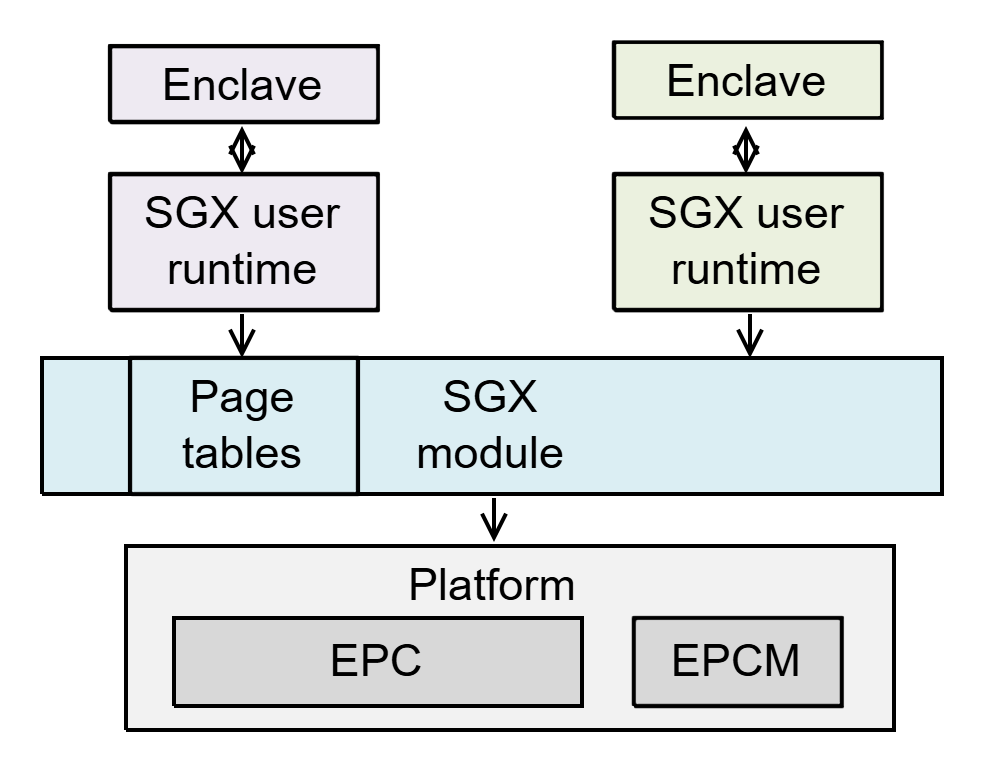}
    \caption{Intel SGX architecture overview.  Sensitive code and data execute inside hardware-protected enclaves within the encrypted EPC, isolated from the untrusted OS and runtime environment to preserve confidentiality and integrity.}
    \label{fig:sgx_arch}
\end{figure}

Intel SGX is a hardware-assisted security extension that provides strong confidentiality and integrity guarantees for code and data by isolating them within a protected memory region known as an enclave. As shown in Figure \ref{fig:sgx_arch}, the code and data inside the enclave are stored in a memory region known as the Enclave Page Cache (EPC). Although the EPC is part of the main memory, it is encrypted and access-controlled by SGX hardware mechanisms within the CPU, ensuring confidentiality and integrity during execution.  

\begin{figure}
    \centering
    \includegraphics[width=.85\linewidth]{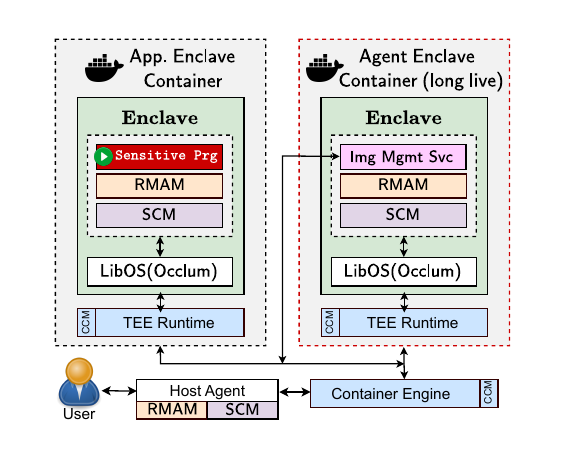}
    \caption{Intel SGX–based \SYSNAME~architecture. 
An \emph{App Enclave} executes sensitive programs, and a persistent \emph{Agent Enclave} manages image deployment, attestation (RMAM), and secure host communication (SCM). 
Together with the host agent and container engine, they realize lightweight per-container TEEs with minimal TCB.}
    \label{fig:shelter_sgx}
\end{figure}

Figure \ref{fig:shelter_sgx} illustrates an implementation of the \SYSNAME~architecture on Intel SGX, which is built upon the CoCo-based enclave-cc runtime\cite{enclave-cc}\footnote{The enclave-cc runtime for CoCo is an open-source project: \url{https://github.com/confidential-containers/enclave-cc}, designed by the 5\textsuperscript{th} author, Jia Zhang.}. \SYSNAME~adheres to the core design principle of constructing a TEE within a container by dividing the enclave space into two distinct types: the Agent Enclave and the App Enclave, thereby enabling fine-grained isolation within the confidential execution environment. The Agent Enclave is responsible for interacting with the host environment and external systems, including security management and resource coordination. In contrast, the App Enclave is dedicated to executing sensitive application logic, ensuring the confidentiality of application code and data. This division effectively minimizes the TCB. In the following sections, we elaborate on the implementation details of each key component within this architecture. 

\begin{figure}
    \centering
    \includegraphics[width=1\linewidth]{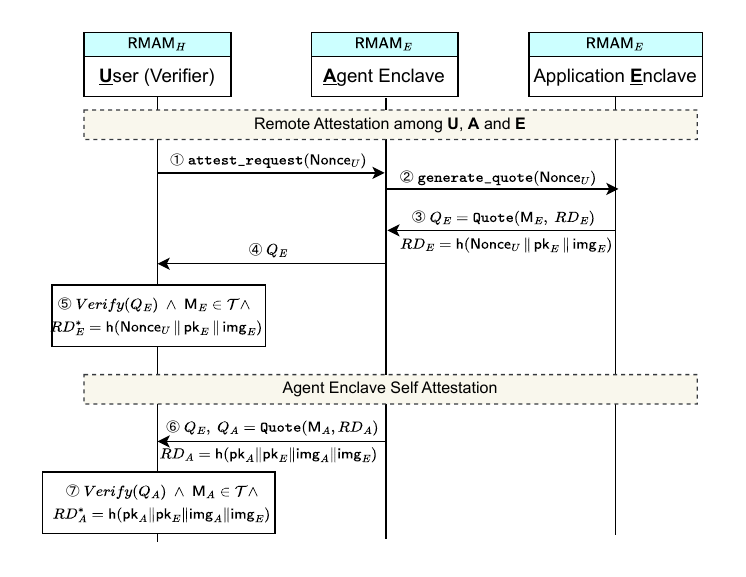}
    \caption{Remote attestation protocol among a verifier \textbf{U}ser, an \textbf{A}gent Enclave and an Application \textbf{E}nclave}
    \label{fig:sgx_RA_protocol}
\end{figure}

\begin{enumerate}
    \item CCM: As shown in Figure \ref{fig:shelter_sgx}, the CCM comprises two core components: the \emph{Container Engine} and the \emph{TEE Runtime}. The Container Engine orchestrates containerized workloads using mainstream container infrastructure such as Docker, while the TEE Runtime manages the enclave lifecycle via Intel SGX instructions. These two components operate in concert to efficiently manage enclave-backed confidential containers. However, since Intel SGX enclaves do not natively support running unmodified Linux applications, seamless migration of legacy workloads is hindered. To bridge this compatibility gap, CCM integrates a library operating system (LibOS) as a compatibility layer that enables secure and transparent execution of unmodified applications within enclaves. To this end, we design and implement Occlum, a lightweight, SGX-optimized LibOS that provides strong POSIX compatibility while ensuring both security and performance for enclave workloads.
    
    \item SCM: As described in section \ref{subsec:sec_comm}, the SCM module is responsible for establishing a secure communication channel between the sensitive program running inside the TEE and the user application in the host (REE side). Firstly, SCM under Intel SGX leverages Occlum's host resource mapping mechanism to construct a data exchange path between the host and the enclave. The data input to the sensitive program is passed to the enclave via a mounted host file system, while the output data is encapsulated via dedicated device files (e.g., /dev/host\_stdout and /dev/host\_stderr), which are redirected back to the host. In this case, the Container Engine acts as an intermediary, capturing and forwarding input and output data. 
    
    Then, the AES-based \emph{Authenticated Encryption with Association Data (AEAD)} algorithm is adopted to ensure confidentiality and integrity during data transmission. Equation \ref{equ:sgx_aead_gen_cipher} shows the calculation of the ciphered data with the related authentication code. $\mathsf{E}_{\mathsf{K}_s}(x)$ denotes the symmetric encryption function with the session key $\mathsf{K}_s$ (E.g., in our implementation, AES-128 with GCM mode is adopted); $\mathsf{HMAC}(x)$ denotes the \emph{Hash-based Message Authentication Code} function. The session key $K_s$ can be generated via equation \ref{equ:sgx_aead_key_gen}, in which $\mathsf{f}_{KDF}(x)$ can be any Key Derivation Function (KDF), $\mathsf{Z}$ is the shared secret between the enclave and the user application (key-agreement protocols, such as ECDH can achieve this), and $\mathsf{MRENCLAVE}_k$ is the measurement of enclave $k$ (E.g., the Agent Enclave or Application Enclave). Equation \ref{equ:sgx_aead_m_gen} reveals how the enclave measurement $\mathsf{MRENCLAVE}_k$ is generated, in which $v^p_i$ denotes the specific hardware/software information, such as the CPU and memory chip serial number, and the essential system configuration files.
    \begin{equation}
        D^*=\{\mathsf{E}_{\mathsf{K}_s}(D)||\mathsf{HMAC}(\mathsf{E}_{\mathsf{K}_s}(D),\mathsf{K}_s)\}
        \label{equ:sgx_aead_gen_cipher}
    \end{equation}
    \begin{equation}
        \mathsf{K}_s=\mathsf{f}_{KDF}(\mathsf{Z}, \mathsf{MRENCLAVE}_k)
        \label{equ:sgx_aead_key_gen}
    \end{equation}
    \begin{equation}
        \mathsf{MRENCLAVE}_k=\mathsf{h}(\mathsf{h}(\mathsf{h}(\mathsf{h}(\mathsf{h}(v^p_1)||v^p_2)||v^p_3)||\ldots)||v^p_m)
        \label{equ:sgx_aead_m_gen}
    \end{equation}
    In our implementation, the trusted application in an enclave utilizes the cryptographic APIs provided by the SGX SDK and OpenSSL. In contrast, the user application uses the OpenSSL library to implement the cryptographic operations described above.  The details on the key management are discussed in section \ref{subsec:key_mgmt}.
    \item RMAM: According to section \ref{subsec:remote_att}, the RMAM is designed to ensure the authenticity and integrity of both sensitive workloads and their corresponding enclaves. To this end, we have designed and implemented a three-party remote attestation protocol, involving a remote \emph{Verifier} (denoted as \textbf{U}), an intermediary \emph{Agent Enclave} (\textbf{A}), and a target \emph{Application Enclave} (\textbf{E}). The Agent Enclave acts as a trusted proxy, facilitating communication and enforcing attestation policies, while the Application Enclave hosts and executes confidential application logic.
    
    Figure \ref{fig:sgx_RA_protocol} shows the protocol, which proceeds as follows. First, the Verifier generates a fresh nonce $\mathsf{Nonce}_U$ and sends it to the Agent Enclave to initiate attestation. The Agent then forwards this nonce to the Application Enclave, requesting a quote:
    \begin{equation}
        Q_E = \mathsf{Quote}(\mathsf{MRENCLAVE}_E, RD_E).
    \end{equation}
    The Application Enclave computes the quote $Q_E$ using the SGX quoting infrastructure, embedding a \texttt{ReportData (RD)} field defined as:
    \begin{equation}
        RD_E = \mathsf{h}(\mathsf{Nonce}_U \,\|\, \mathsf{pk}_E \,\|\, \mathsf{img}_E),
    \end{equation}
    where $\mathsf{pk}_E$ is the public key of enclave $E$, $\mathsf{img}_E$ is the hash of the container image identity and $\mathsf{hash}$ is a collision-resistant hash function (e.g., SHA-256). The quote $Q_E$, which includes the enclave measurement $\mathsf{MRENCLAVE}_E$ ($\mathsf{M}_E$) and the bound report data, is returned to the Agent, who relays it to the Verifier.
    
    Upon receiving $Q_E$, the Verifier performs SGX quote verification using DCAP, checks the integrity and validity of $\mathsf{MRENCLAVE}_E$, and verifies that the embedded report data matches the expected value. This binding ensures that the quote is fresh and originates from an enclave possessing $\mathsf{pk}_E$. The Verifier accepts the attestation only if all checks pass and $\mathsf{MRENCLAVE}_E$ belongs to a predefined trust set $\mathcal{T}$:

    \begin{equation}
        \begin{split}
            \mathsf{Verify}(Q_E) \wedge RD_E = \mathsf{h}(\mathsf{Nonce}_U \,\|\, \mathsf{pk}_E \,\|\, \mathsf{img}_E) \wedge \\
            \mathsf{MRENCLAVE}_E \in \mathcal{T}.
        \end{split}
    \end{equation}
    
    Optionally, the Agent Enclave may also generate its own quote $Q_A$ with
    \begin{align}
        Q_A &= \mathsf{Quote}(\mathsf{MRENCLAVE}_A, RD_A) \\
        RD_A &= \mathsf{h}(\mathsf{pk}_A \| \mathsf{pk}_E \| \mathsf{img}_A \| \mathsf{img}_E)
    \end{align}
    enabling the Verifier to perform transitive trust validation through the Agent. This mechanism supports flexible deployment models where the Verifier cannot directly interact with the Application Enclave but still wishes to establish trust in it.
    
    By embedding both enclave identities and their corresponding container image digests into the report data, this design ensures that the Agent is not only running trusted code ($\mathsf{MRENCLAVE}_A$), but is also intentionally associated with a specific Application Enclave ($\mathsf{pk}_E$) and its execution context ($\mathsf{img}_E$). Moreover, including the Agent’s own image hash, $\mathsf{img}_A$, binds the attestation to a concrete container's provenance, preventing rollback to outdated or malicious container versions. This binding enables end-to-end auditability and significantly strengthens the integrity guarantees of the trust chain in containerized TEE deployments.

\end{enumerate}

\subsection{\SYSNAME ~for Intel TDX and AMD SEV}\label{subsec:imp_tdx}
\begin{figure}
    \centering
    \includegraphics[width=1\linewidth]{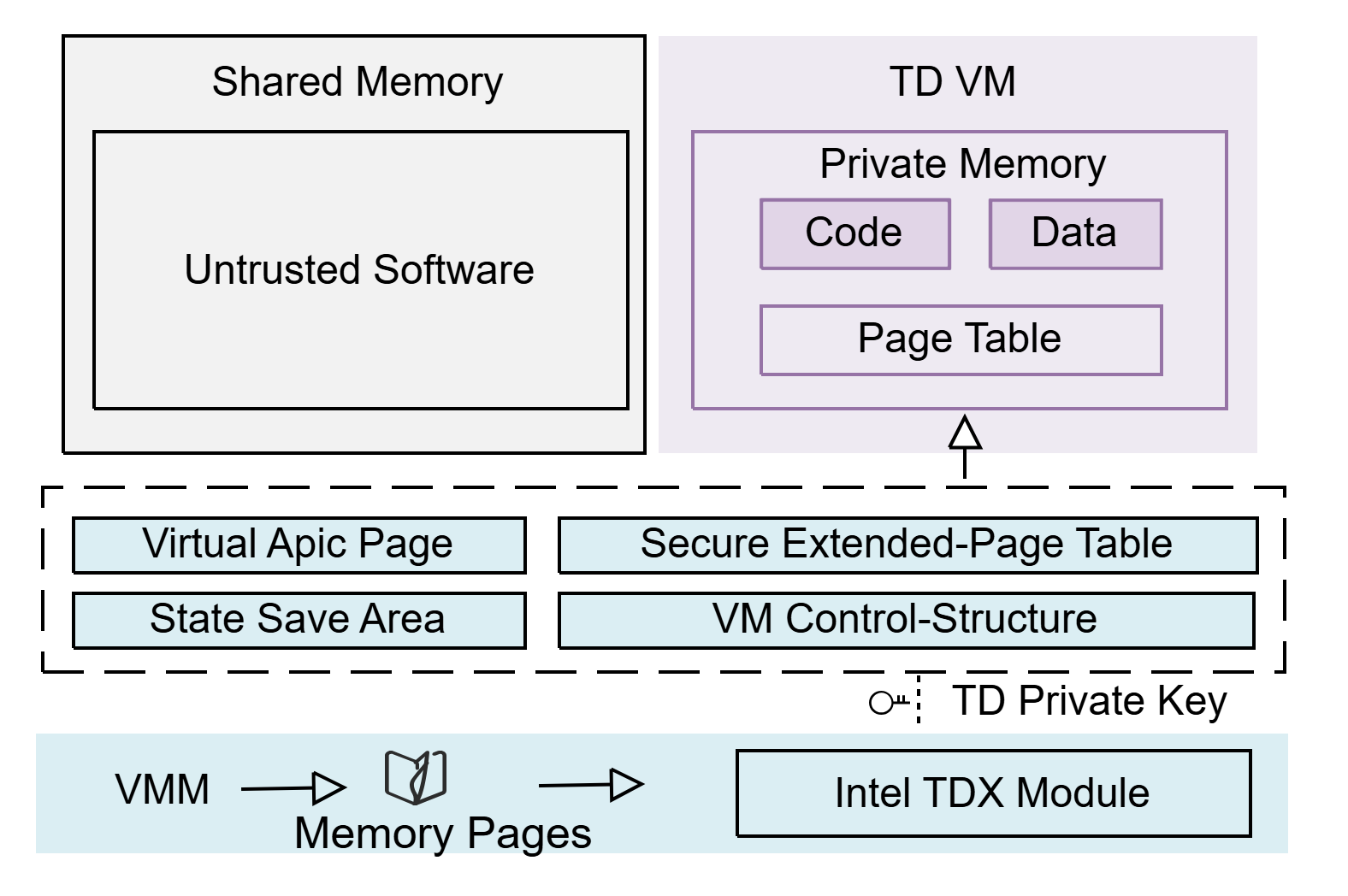}
    \caption{Intel TDX architecture. Each virtual machine runs inside a hardware-isolated \emph{TD} whose memory and state are protected from the host, providing the foundation for container-level confidential computing.}
    \label{fig:tdx_arch}
\end{figure}

Intel TDX is a hardware-based confidential computing technology designed to enhance the security of data and code in virtualized environments. As shown in Figure \ref{fig:tdx_arch}, in the Intel TDX architecture, each VM is mapped to a distinct TD, serving as its isolated execution environment. The TD leverages hardware-enforced encryption and isolation mechanisms to protect the memory and execution state of the TD VM, ensuring that its data remains secure from access or modification by the host operating system and other virtual machines. Additionally, Intel TDX introduces the TDX Module, which manages the creation and lifecycle of TDs while coordinating with CPU security extensions to ensure the integrity and isolation of TD execution. 

\begin{figure}
    \centering
    \includegraphics[width=0.65\linewidth]{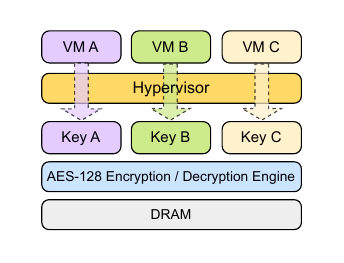}
    \caption{AMD SEV architecture. The PSP encrypts each VM’s memory with a unique key, preventing the hypervisor from accessing plaintext data and enabling hardware-enforced VM isolation.}
    \label{fig:sev_arch}
\end{figure}

AMD SEV employs hardware-based encryption to ensure the confidentiality of VM memory, preventing data from being decrypted even if the memory is physically accessed. As shown in Figure \ref{fig:sev_arch}, each VM’s memory in the AMD SEV architecture is protected using a unique encryption key. These keys are generated and securely managed by the Platform Security Processor (PSP) and are used internally by the CPU to transparently perform memory encryption and decryption at runtime. The encryption keys are never exposed to the host operating system or the hypervisor, thereby guaranteeing the confidentiality of the VM memory.

In summary, both Intel TDX and AMD SEV represent VM-based TEEs. Due to the highly abstract nature of the architecture design, the \SYSNAME{} architecture built on these two platforms remains consistent at the overall design level, with differences only in specific technical implementations. Therefore, this paper presents a unified discussion of the implementation details of \SYSNAME~on both platforms. 

\begin{figure}
    \centering
    \includegraphics[width=.65\linewidth]{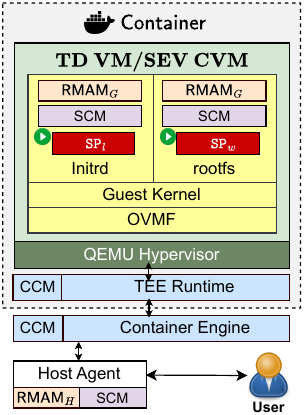}
    \caption{\SYSNAME~architecture on Intel TDX and AMD SEV. Each container hosts a confidential VM (TD VM/SEV CVM) for sensitive workloads, while management components remain outside the trusted boundary, achieving a unified lightweight TCB across heterogeneous VM-based TEEs.}
    \label{fig:shelter_tdx_sev}
\end{figure}

Figure \ref{fig:shelter_tdx_sev} presents an instance of implementing the \SYSNAME~architecture on Intel TDX/AMD SEV, extending its security model from process-level to VM-level isolation. Following \SYSNAME~’s core design principle, we establish a TD VM/CVM within the container. Instead of migrating the entire container and its management engine into the TD VM/CVM, which would unnecessarily inflate the TCB, \SYSNAME~’s TD VM/CVM is dedicated to executing runtime-protected code. The following sections provide detailed explanations of the implementation of each key component in this architecture.

\begin{enumerate}
    \item CCM:As shown in Figure \ref{fig:shelter_tdx_sev}, this module relies on two crucial components: the \emph{Container Engine} and the \emph{TEE Runtime}. The Container Engine is built upon the mainstream container technology Docker. At the same time, the TEE Runtime is implemented on QEMU platforms that support Intel TDX and AMD SEV, respectively, on different physical machines. The right part of Figure \ref{fig:TDX_SEV_CCM_SCM} shows the process of how a CVM is launched. Upon receiving the user's requests, \emph{Host Agent} will send commands to the Docker engine to spawn a new container, in which the VMM \emph{QEMU} will run, and the confidential virtual machine is finally launched.
    
    To provide a streamlined and efficient guest environment, we apply lightweight customization to the essential system components required for launching TD VMs or SEV CVMs, including the \emph{Guest Kernel}, the initial RAM disk \emph{Initrd}, and the virtual machine firmware Open Virtual Machine Firmware (\emph{OVMF}). 
    OVMF is responsible for VM booting, which initializes the virtualized hardware environment and loads the \emph{Guest Kernel} and \emph{Initrd} image into the protected memory region. Then, the \emph{Guest Kernel} mounts the \emph{Initrd} as a temporary root filesystem to execute initialization tasks, including some lightweight sensitive programs ($\mathsf{SP}_l$ in \emph{Initrd} in figure \ref{fig:TDX_SEV_CCM_SCM}). For long-term tasks, \emph{Initrd} loads the user-provided root file system (rootfs), which is typically encrypted and contains the long-term sensitive tasks. Once the user rootfs is successfully mounted, \emph{Initrd} will be discarded and control will be handed over to the actual rootfs ($\mathsf{SP}_w$ in \emph{Initrd} in figure \ref{fig:TDX_SEV_CCM_SCM}).
   
    \item SCM: 
\begin{figure}
    \centering
    \includegraphics[width=0.8\linewidth]{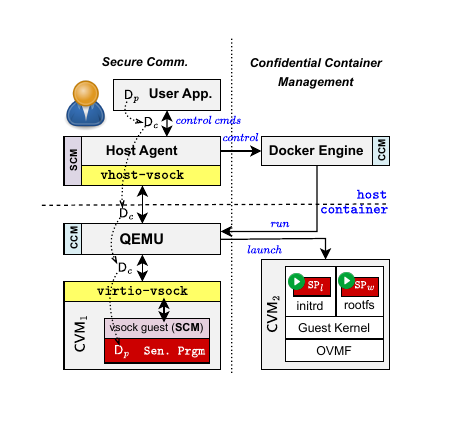}
    \caption{Illustration of CCM and SCM on Intel TDX and AMD SEV. The right part shows how the Host Agent cooperates with Docker Engine to control QEMU and launch a confidential VM (CVM\textsubscript{2}) that contains essential boot components (\textit{OVMF}, \textit{Initrd}, and \textit{Guest Kernel}) and image management services.
The left part depicts the SCM pathway, where user data \(D_p\) is encrypted into \(D_c\) and transmitted via the attested \textit{vhost-vsock}–QEMU–\textit{virtio-vsock} channel, ensuring end-to-end confidentiality between the host and the sensitive program inside CVM\textsubscript{1}.}
    \label{fig:TDX_SEV_CCM_SCM}
\end{figure}
Shown in the left part of Figure~\ref{fig:TDX_SEV_CCM_SCM}, the SCM establishes a secure data path between the host environment and sensitive programs running inside a TEE, implemented as a TD VM (under Intel TDX) or a CVM (under AMD SEV). To support low-latency host-guest communication, the SCM utilizes the Virtual Socket (vsock) transport, which enables direct data exchange between the host and the guest VM, eliminating the need for conventional networking stacks such as TCP/IP. This design is platform-agnostic: switching between TDX and SEV requires only modifications to the container's launch configuration.

The \emph{Host Agent} utilizes the \texttt{vhost-vsock} interface (i.e., \texttt{/dev/vhost-vsock}), a kernel module provided by the host operating system, which operates as the backend for vsock communication. On the VM side, QEMU exposes a \texttt{virtio-vsock} device to the guest, which serves as the frontend vsock interface. This setup allows data packets to be transferred entirely within kernel space, avoiding costly transitions between user space and kernel space. Inside the confidential virtual machine, we implement a \emph{vsock guest} service that interfaces directly with the \texttt{virtio-vsock} device, enabling sensitive applications to participate in the vsock-based communication channel transparently.

Crucially, to ensure the confidentiality and integrity of transmitted data—even across potentially untrusted host components—the vsock communication channel is secured using AEAD. Equation~\ref{equ:sgx_aead_gen_cipher} illustrates the encryption process.
Each data packet from the vsock host (\emph{Host Agent}) to the vsock guest is encrypted using AES-GCM, producing both a ciphertext and an authentication tag. As shown in the figure, user data $\mathsf{D}_p$, which is in plaintext, will be encrypted ($\mathsf{D}_c$) before being sent to the untrusted components, such as the \emph{Host Agent}, the container, and QEMU within. Once $\mathsf{D}_c$ is received by the CVM, the sensitive program within it will restore the data to plaintext for further processing (the arrows with dashed lines illustrate such a procedure).
The same mechanism is applied in reverse for response messages. Both the user application and sensitive program in CVM use OpenSSL to perform cryptographic operations and authentication checks.

Additionally, the encrypted messages are tied to the attested identity of the TD VM or CVM. This ensures that the host communicates only with a verified and trusted instance of the sensitive program. Details of the key derivation and distribution process are presented in Section~\ref{subsec:key_mgmt}.
    
    \item RMAM: In TDX/SEV-based implementations, \SYSNAME~leverages the inherent remote attestation (RA) mechanisms of TDX/SEV to assess the authenticity of CVMs (TD VM or CVM) as well as the containers they are located on. We will discuss the remote attestation protocols under TDX and SEV, respectively.
    \begin{figure}
    \centering
    \includegraphics[width=1\linewidth]{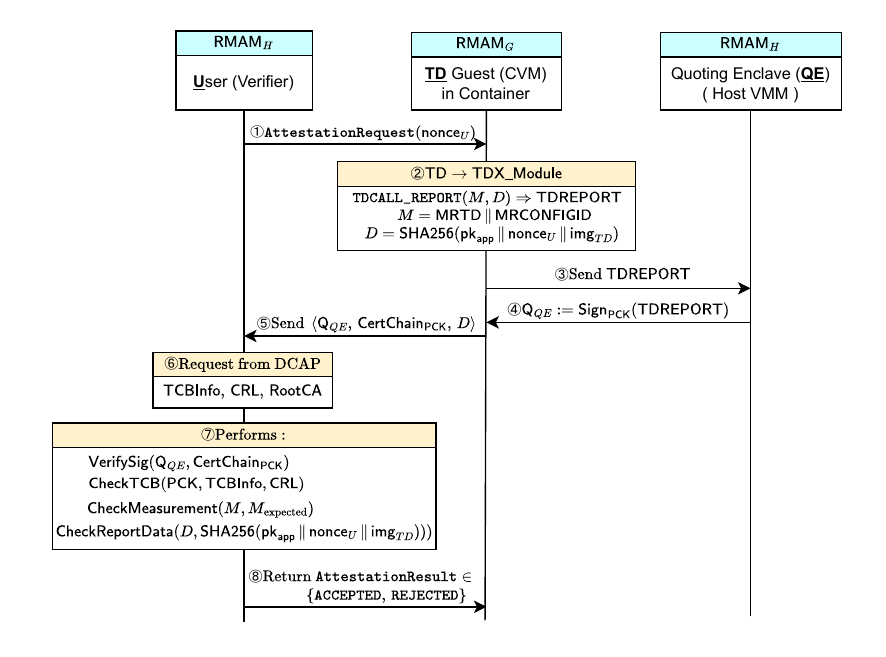}
    \caption{Illustration of the TDX-based remote attestation protocol.}
    \label{fig:tdx_RA_protocol}
    \end{figure}

    $\bullet$~~\emph{RA with TDX:} Figure~\ref{fig:tdx_RA_protocol} illustrates the TDX-based remote attestation protocol, which involves three primary entities: a Verifier (User), a TD Guest (CVM), and the Quoting Enclave (QE). To accommodate the distinct responsibilities of the host and the TD Guest, we introduce two types of Remote Measurement and Attestation Modules (RMAM): the host-side RMAM ($\mathsf{RMAM}_H$), which handles attestation-related operations on the host, and the guest-side RMAM ($\mathsf{RMAM}_{G}$), which operates within the confidential VM to perform in-guest attestation tasks. 
    
    The remote attestation process begins with the Verifier sending an $\mathsf{AttestationRequest}$ containing a randomly generated nonce to the TD. The nonce ensures attestation freshness and prevents replay attacks by binding the attestation to a unique challenge. Upon receiving the request, the TD generates a $\mathsf{TDREPORT}$ using the $\texttt{TDCALL\_REPORT}$ instruction from the TDX Module. The report contains a measurement value $M$ and a $\mathsf{ReportData}$ $D$, calculated as $\mathsf{SHA256}(\mathsf{pk}_{app} \| \mathsf{nonce}_U \| \mathsf{img}_{TD})$, linking the TD’s public key, container image hash ($\mathsf{img}_{TD}$) to the nonce. Then, the TD sends the $\mathsf{TDREPORT}$ to the Quoting Enclave (QE), which signs it with the Platform Certification Key (PCK) to produce a quote ($\mathsf{Q}_{QE}$). The TD then returns the $\mathsf{Q}_{QE}$, the $\mathsf{CertChain_PCK}$, and $\mathsf{ReportData}$ to the Verifier, which uses these to validate the TD’s attestation. After receiving these data, the Verifier fetches platform metadata from the DCAP service, including TCB Information and CRLs, and performs several checks: verifying the $\mathsf{Q}_{QE}$ signature, validating the certificate chain, confirming the TCB status, comparing the measurement $M$, and ensuring the $\mathsf{ReportData}$ matches $\mathsf{SHA256}(\mathsf{pk}_{app} || \mathsf{nonce}_U \| \mathsf{img}_{TD})$. If all checks pass, the Verifier returns an \texttt{ACCEPTED} result, allowing the TD to proceed with secure operations. If any check fails, the attestation is \texttt{REJECTED}, and the TD cannot continue.
    
    This process ensures that the TD is running in a trusted environment and that the attestation data is fresh, valid, and secure against replay attacks.

    \begin{figure*}[htbp]
      \centering
      \includegraphics[width=0.9\textwidth]{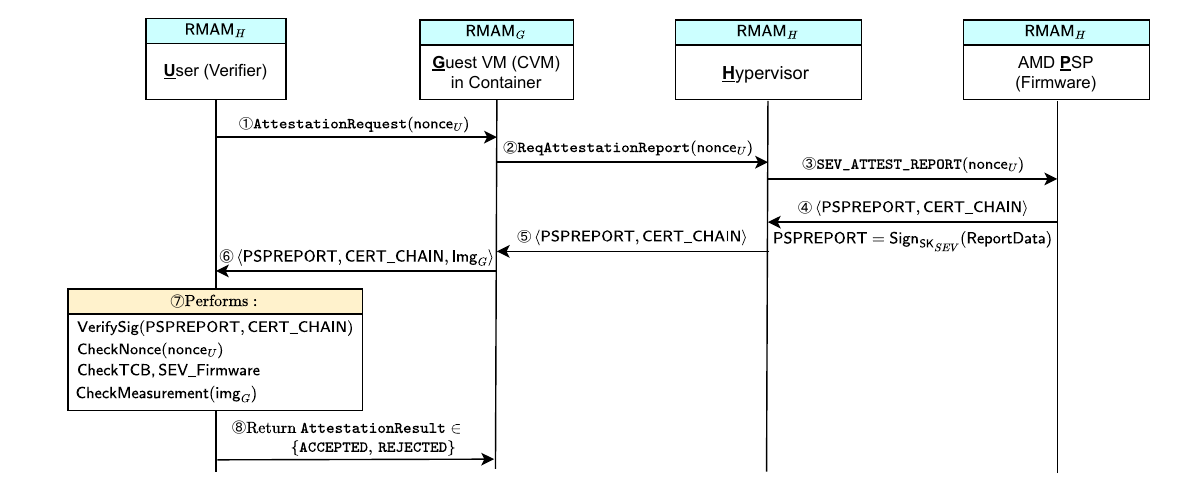}
      \caption{Illustration of the SEV-based remote attestation protocol.}
      \label{fig:sev_RA_protocol}
    \end{figure*}
    $\bullet$~~\emph{RA with SEV:} Although the SEV and TDX attestation protocols follow a broadly similar architecture, their low-level operations,  such as report generation, isolation primitives, and certificate handling, reflect distinct design choices rooted in platform-specific trust models. 
    
    To assess the trustworthiness of a confidential virtual machine (Guest VM, GVM) protected by AMD SEV, a remote attestation protocol is executed as illustrated in Figure~\ref{fig:sev_RA_protocol}. The protocol begins when the verifier issues an $\texttt{AttestationRequest}(\mathsf{nonce}_U)$ to the GVM (step \textcircled{1}), where the nonce ensures freshness of the attestation. The GVM forwards this request to the Hypervisor, which invokes the PSP firmware via the $\mathsf{SEV\_ATTEST\_REPORT}(\mathsf{nonce}_U)$ command (steps \textcircled{2}-\textcircled{3}). 
    
    Then, the PSP constructs an attestation report $\mathsf{PSPREPORT}$. Internally, it computes a SHA-256 measurement over the GVM’s initial memory pages defined at launch, producing the measurement field. This reflects the cryptographic fingerprint of the VM’s initial state, including the bootloader, kernel, and configuration data. The PSP also incorporates the nonce for freshness, the guest-provided $\mathsf{ReportData}$ (often set to a digest of application metadata, such as $\mathsf{SHA256}(\mathsf{pk}_{app} \,\|\,\mathsf{container}_{id}))$, and metadata such as Virtual Machine Privilege Level (VMPL), firmware version, and TCB levels. The VMPL field in the attestation report reveals the VM privilege level at which the confidential guest is executing. This enables the verifier to assess whether sensitive code runs within a sufficiently privileged and isolated context (e.g., VMPL0, most privileged; VMPL3, least privileged). By incorporating VMPL, SEV binds the measurement not only to the initial image but also to its runtime trust domain, preventing misconfiguration or privilege escalation within the guest.
    
    The complete report is then digitally signed using the platform’s attestation private key ($\mathsf{sk}_{SEV}$), which resides within the AMD Secure Processor. To enable verification, the PSP provides a certificate chain ($\mathsf{CERT\_CHAIN}$) rooted at AMD’s hardware root of trust (steps \textcircled{4}–\textcircled{5}). This chain consists of the Platform Endorsement Key (PEK) certificate, which is signed by the Chip Endorsement Key (CEK), and the CEK is, in turn, signed by AMD’s root certificate authority. The verifier will validate the report's signature using this chain and verify that the CEK/PEK remain within a trusted TCB, using AMD’s published revocation lists and TCB metadata.
    
    The GVM optionally attaches a container image hash, a SHA-256 digest of the expected VM launch image ($\mathsf{img}_G$), and forwards the signed report and certificates to the verifier (step \textcircled{6}). Upon receipt, the verifier performs a series of checks (step \textcircled{7}): it verifies the report’s signature ($\mathsf{VerifySig}(\mathsf{PSPREPORT}, \mathsf{CERT\_CHAIN})$), confirms that the embedded nonce matches the challenge, validates the certificate chain against AMD’s root CA, and compares the reported measurement with a locally stored reference hash (e.g., $\mathsf{CheckMeasurement}(\mathsf{img}_G)$). If all validations succeed, an attestation result is returned to the GVM (step \textcircled{8}), indicating acceptance or rejection.
\end{enumerate}

\section{Performance Evaluation}\label{sec:perf_eval}
\subsection{Experimental Setup}\label{subsec:exp_setup}

\begin{table*}[t]
\caption{Configuration of the selected platforms for experiments.}
\label{tbl:exp_platform_config}
\scriptsize
\centering
\begin{tabular*}{\linewidth}{l@{\extracolsep{\fill}}ll}
\hline
Platform & Hardware Configuration & OS \& Kernel \\
\hline
Intel SGX/TDX &
\begin{tabular}[c]{@{}l@{}}
CPU: Intel(R) Xeon(R) Platinum 8558 48-Core@2.1GHz \\
RAM: 256GB DDR5 4800MT/s \\
HDD: Samsung SSD 990 EVO Plus 2TB
\end{tabular} &
Ubuntu 24.04.2 LTS with kernel 6.8.0-1028-intel \\[4pt]

AMD SEV &
\begin{tabular}[c]{@{}l@{}}
CPU: AMD EPYC 7401 24-Core@2.0GHz \\
RAM: 64GB DDR4 2400MT/s \\
HDD: Crucial SSD BX500 1TB
\end{tabular} &
Ubuntu 24.04.1 LTS with kernel 6.8.0-63-generic \\[4pt]

CoCo &
\begin{tabular}[c]{@{}l@{}}
Same as the compared platform hardware configuration
\end{tabular} &
\begin{tabular}[c]{@{}l@{}}
Same as the compared platform OS
\end{tabular} \\
\hline
\end{tabular*}
\end{table*}

We have implemented \SYSNAME~based on two representative hardware platforms: one supporting Intel SGX and TDX, and the other supporting AMD SEV. To evaluate the performance differences between \SYSNAME~and CoCo (one of the most representative confidential container projects), we deployed CoCo on the same platforms under identical experimental conditions. Table \ref{tbl:exp_platform_config} presents the configuration of the selected platforms used in our experiments.

Our experiments evaluate \SYSNAME~from two main performance perspectives and one security aspect. First, for Intel SGX, we use \texttt{sysbench-cpu}, \texttt{sysbench-threads}, \texttt{fio-IOPS}, and \texttt{iperf3} to demonstrate the performance of \SYSNAME~compared to the baseline SGX SDK. Second, for Intel TDX and AMD SEV, we use the open-source Unixbench\cite{Unixbench} to compare \SYSNAME~with CoCo, highlighting performance differences across these TEE-supported platforms. Finally, we discussed \SYSNAME's security performance.

\subsection{Comprehensive Performance Benchmarking  }\label{subsec:compre_test}
To evaluate the performance advantages of our proposed \SYSNAME~ framework over existing TEE-based container systems, we conducted a series of benchmark experiments across three representative cloud-native TEEs: Intel SGX, Intel TDX, and AMD SEV.
For Intel TDX and AMD SEV, we used the widely used \texttt{UnixBench} suite to assess computational performance via system calls, process creation, context switching, and arithmetic throughput for both integer and floating-point workloads.
For Intel SGX, however, only a subset of UnixBench-compatible workloads can be executed due to its limited EPC memory and restricted access to kernel-level operations. Therefore, we used \texttt{sysbench} and \texttt{FIO} to measure CPU computation, multithread scalability, and I/O throughput, which together provide a representative view of SGX performance under enclave constraints.
This configuration enables a fair, fine-grained evaluation of \SYSNAME’s performance across different TEE architectures and runtime environments.

\begin{figure*}
    \centering
    \includegraphics[width=0.75\linewidth]{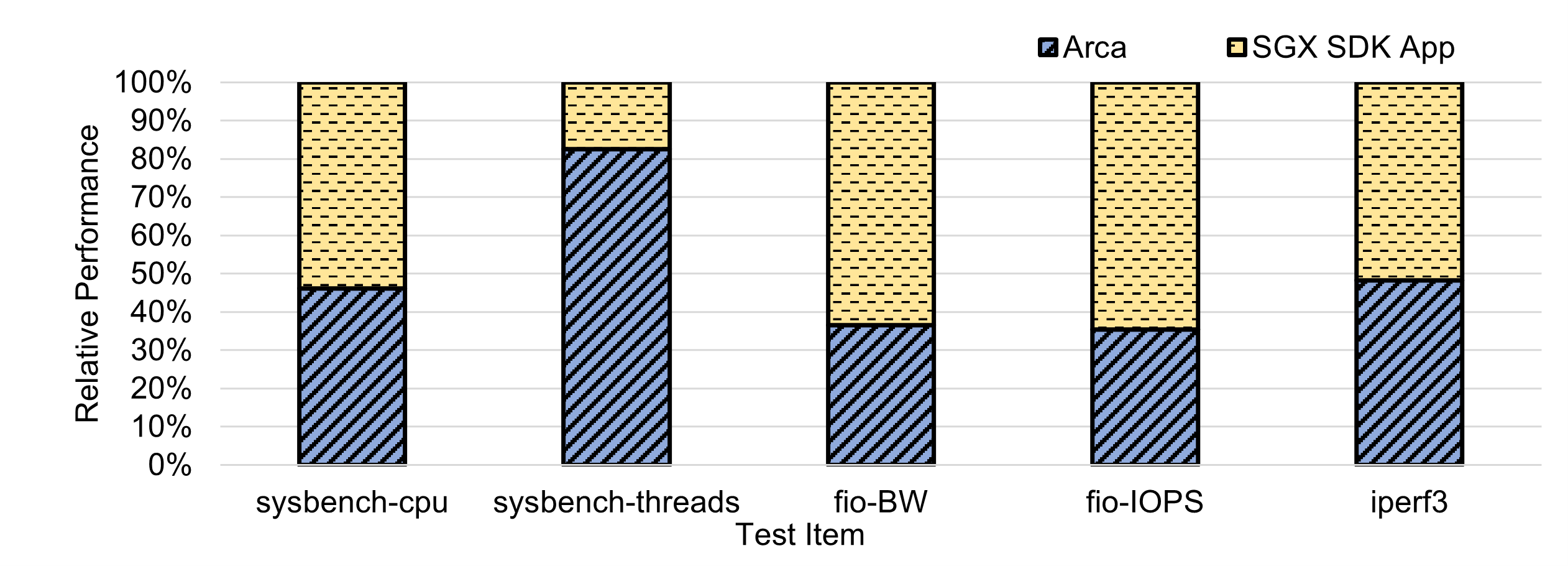}
    \caption{Performance Comparison of \SYSNAME~and Intel SGX SDK on Intel SGX. \SYSNAME~maintains near-native performance on Intel SGX while delivering stronger isolation and better integration with containerized environments than the native SGX SDK}
    \label{fig:test_sgx}
\end{figure*}

\subsubsection{Performance Evaluation on Intel SGX Platform}
Figure~\ref{fig:test_sgx} shows the comparative results between \SYSNAME~and native SGX SDK applications on the SGX platform, evaluated using \texttt{sysbench-cpu}, \texttt{sysbench-threads}, \texttt{fio-BW}, \texttt{fio-IOPS}, and \texttt{iperf3}. 
Overall, \SYSNAME~achieves comparable or superior performance in CPU- and thread-intensive workloads, while exhibiting notable degradation in I/O-bound tests, particularly in \texttt{fio-BW} and \texttt{fio-IOPS}. 
The following analysis explains these observations in detail.

a) \emph{I/O-intensive workloads:} 
In the \texttt{fio-BW} and \texttt{fio-IOPS} benchmarks, \SYSNAME~demonstrates a substantial performance drop relative to the native SGX SDK. 
This degradation primarily results from the multi-layered I/O path introduced by the containerized TEE abstraction. 
Unlike SGX SDK applications, which can issue system calls via a relatively direct \texttt{OCALL} interface, \SYSNAME~routes I/O operations through additional \emph{shim} and \emph{proxy layers} that bridge the container runtime and the host environment. 
Each operation therefore crosses multiple trust boundaries—container runtime, enclave boundary, and host kernel—introducing extra data copies, boundary validations, and context switches.

The limited EPC further exacerbates this issue. 
To guarantee confidentiality, \SYSNAME~enforces strict buffer isolation, duplicating I/O data between trusted and untrusted memory regions. 
This duplication causes frequent EPC paging, encryption/decryption operations, and TLB flushes, all of which degrade I/O throughput. 
In contrast, SGX SDK applications can directly map untrusted memory for I/O buffers, avoiding these redundant copies. 
Hence, the combination of multi-layer communication, extra context transitions, and conservative buffer management explains the sharp performance decline in I/O-heavy scenarios.

b) \emph{Thread-intensive workloads: }
Conversely, \SYSNAME~outperforms the SGX SDK baseline in the \texttt{sysbench-threads} test, indicating better scalability under concurrent workloads. 
This advantage stems from \SYSNAME~'s lightweight threading model and container-aware scheduling mechanism. 
The SGX SDK statically allocates a limited number of Thread Control Structures (TCS) at enclave creation, with each \texttt{EENTER}/\texttt{EEXIT} transition incurring high latency. 
By contrast, \SYSNAME~multiplexes multiple user-level threads across a smaller pool of enclave threads, effectively reducing enclave transitions and improving CPU utilization.

Additionally, \SYSNAME~leverages the container runtime’s cgroup-based scheduling and CPU affinity configurations to distribute threads more evenly across physical cores. 
Synchronization between threads is also optimized through shared in-container memory channels, eliminating the need for frequent enclave–host event signaling. 
These enhancements yield superior parallelism and reduced synchronization overhead, allowing \SYSNAME~to achieve higher throughput in thread-heavy benchmarks despite its additional abstraction layer.

\emph{Summary: }
Although \SYSNAME~shows moderate performance degradation in some I/O-intensive benchmarks compared with the native SGX SDK, this does not affect its practical usability. More importantly, \SYSNAME~provides several architectural advantages that make it a stronger foundation for trusted application deployment.
It provides a unified, container-native execution environment that simplifies enclave deployment and management via standard container interfaces. It enforces stronger runtime isolation and attestation between trusted and untrusted components, enhancing system security in multi-tenant scenarios. In addition, its seamless integration with container ecosystems (e.g., containerd, Kubernetes) enables enclave-protected workloads to join cloud-native pipelines with minimal modification. Finally, its modular abstraction design facilitates portability and maintainability, supporting future extensions to TEEs such as TDX and SEV without major code changes.

In summary, while \SYSNAME~sacrifices a small portion of raw performance, it achieves significantly better usability, scalability, and security flexibility than conventional SGX SDK applications.

\begin{figure*}
    \centering
    \includegraphics[width=1\linewidth]{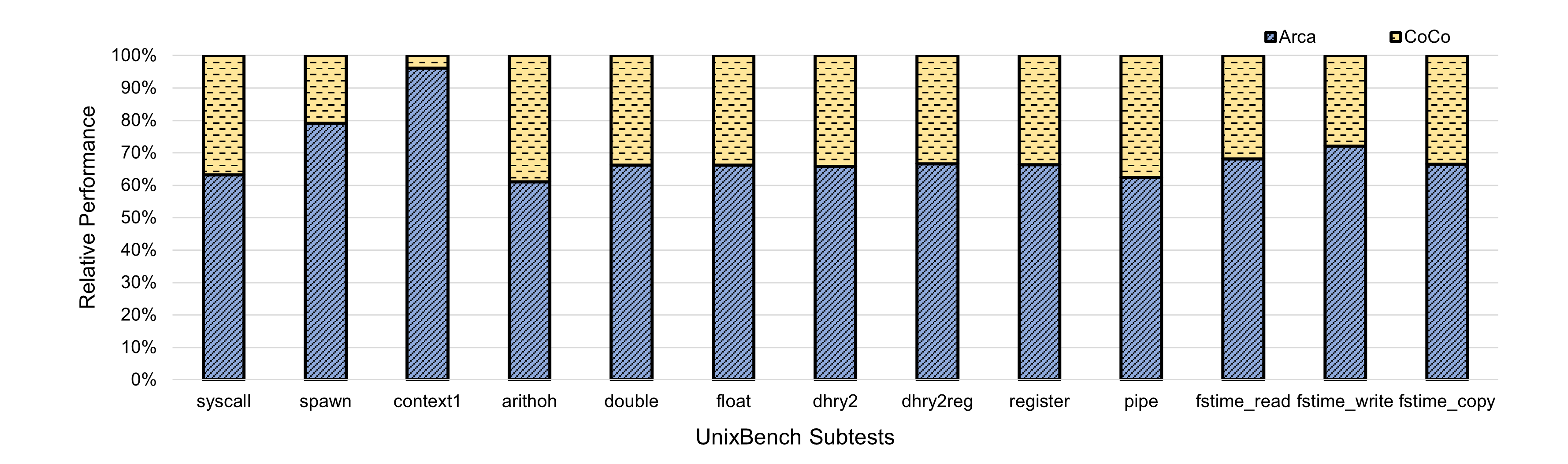}
    \caption{UnixBench performance comparison between \SYSNAME~and CoCo on Intel TDX. \SYSNAME~consistently surpasses CoCo across nearly all subtests, with a pronounced improvement in \texttt{context1} and steady gains in the \texttt{fstime-*} group.
The smaller margin in \texttt{pipe} reflects that inter-process communication benefits less from the flattened TEE-in-Container execution path.}
    \label{fig:test_tdx}
\end{figure*}

\subsubsection{Performance Evaluation on Intel TDX Platform}

Figure~\ref{fig:test_tdx} compares the performance of \SYSNAME~ and CoCo on the Intel TDX platform. Across most UnixBench subtests, \SYSNAME~shows comparable or improved performance, indicating that the TEE-in-Container architecture works efficiently with TDX’s virtualized trusted domain. Below, we analyze the dominant performance trends reflected in the results.

\textit{a) Execution efficiency: }
In CPU- and syscall-intensive benchmarks, including \texttt{context1}, \texttt{syscall}, \texttt{arithoh}, \texttt{dhry2}, and \texttt{register}, \SYSNAME~consistently delivers distinct better performance than CoCo. The most noticeable gain occurs in \texttt{context1}, which stresses frequent context switching. This improvement is attributed to the fact that, in the TEE-in-Container design of \SYSNAME, the TD~VM kernel serves as the sole trusted scheduling domain. As a result, context-switch decisions are executed directly inside the TD~VM without requiring an additional mediation layer from a container runtime, thereby reducing boundary transitions and scheduling latency.

A similar performance advantage is observed in the \texttt{syscall} benchmark, where applications in \SYSNAME~issue system calls that interact directly with the trusted kernel within the TD~VM. In contrast, CoCo retains container-level namespace isolation and security mediation within the VM, causing system calls to traverse an extra validation stage before reaching the kernel. This longer syscall pipeline introduces additional context-switch overhead, resulting in higher latency.

Other compute-oriented microbenchmarks such as \texttt{arithoh}, \texttt{dhry2}, and \texttt{register} follow the same trend. Since these workloads primarily rely on CPU execution within the trusted kernel, the flattened software hierarchy of \SYSNAME~preserves more native performance, while CoCo’s container-level mediation slightly restricts throughput. Collectively, these results demonstrate that reducing the number of mediator components within the trusted execution boundary improves CPU utilization and kernel interaction efficiency under TDX.

\textit{b) Creation responsiveness: }
In the \texttt{spawn} benchmark, which evaluates the responsiveness of process creation and activation, \SYSNAME~achieves a clear performance advantage over CoCo. This test stresses the cost of constructing a new trusted execution context. In CoCo, this procedure requires assembling container-level isolation metadata within the TD~VM before the kernel can mark the new process as runnable. In contrast, the TEE-in-Container design in \SYSNAME~aligns the trusted boundary directly with the TD~VM kernel, allowing process instantiation to be completed entirely within one trusted scheduling and resource-management domain. Eliminating the redundant container-runtime mediation shortens the initialization path and reduces both kernel coordination and state propagation overhead, thus enabling faster process creation.

A similar performance pattern is reflected in file-operation benchmarks such as \texttt{fstime\_write} and \texttt{fstime\_copy}, which involve frequent metadata updates and buffer allocation. Since system calls issued by \SYSNAME~applications are handled directly by the trusted TD~VM kernel, filesystem operations avoid the extra namespace and policy checks required in CoCo’s container-in-TEE model. This direct interaction further improves the efficiency of context activation and state synchronization during I/O-intensive initialization phases. Therefore, the enhanced creation responsiveness observed in these tests confirms that reducing the number of isolation layers within the TEE effectively accelerates the formation of trusted execution contexts under TDX.

\textit{c) Synchronization neutrality: }
For benchmarks such as \texttt{pipe}, \texttt{float}, \texttt{double}, and \texttt{dhry2reg}, \SYSNAME~and CoCo exhibit nearly identical performance. These tests primarily depend on in-guest scheduling efficiency or on hardware execution pipelines such as the floating-point unit. Since TDX enforces trusted isolation at the VM granularity, inter-process communication and arithmetic execution proceed without triggering heavy trust-boundary transitions. As a result, the container boundary in \SYSNAME~introduces negligible synchronization overhead in these scenarios, leading to performance parity. The slight disadvantage observed in the \texttt{pipe} test reflects a minimal coordination cost when two communicating entities reside across the container--TD~VM interface, but the difference remains marginal and does not alter the overall trend.

\textit{Summary: }
\SYSNAME~consistently matches or exceeds CoCo’s performance across representative UnixBench workloads on the TDX platform. Notable advantages are observed in \texttt{context1}, \texttt{syscall}, and \texttt{spawn}, where the TEE-in-Container execution boundary reduces system-call mediation and eliminates redundant isolation-layer coordination during context activation. File-operation tests such as \texttt{fstime\_write} and \texttt{fstime\_copy} further confirm that maintaining a single trusted kernel environment benefits state synchronization and metadata management. Meanwhile, compute-bound and communication-heavy workloads yield near-identical results, demonstrating that TDX’s VM-centric trust model effectively neutralizes the residual impact of the container layer. These findings indicate that aligning trusted execution with the VM as the primary isolation domain enables \SYSNAME~to improve runtime efficiency while preserving the strong security guarantees of hardware-backed virtualization.

\subsubsection{Performance Evaluation on Intel SEV Platform}
\begin{figure*}
    \centering
    \includegraphics[width=1\linewidth]{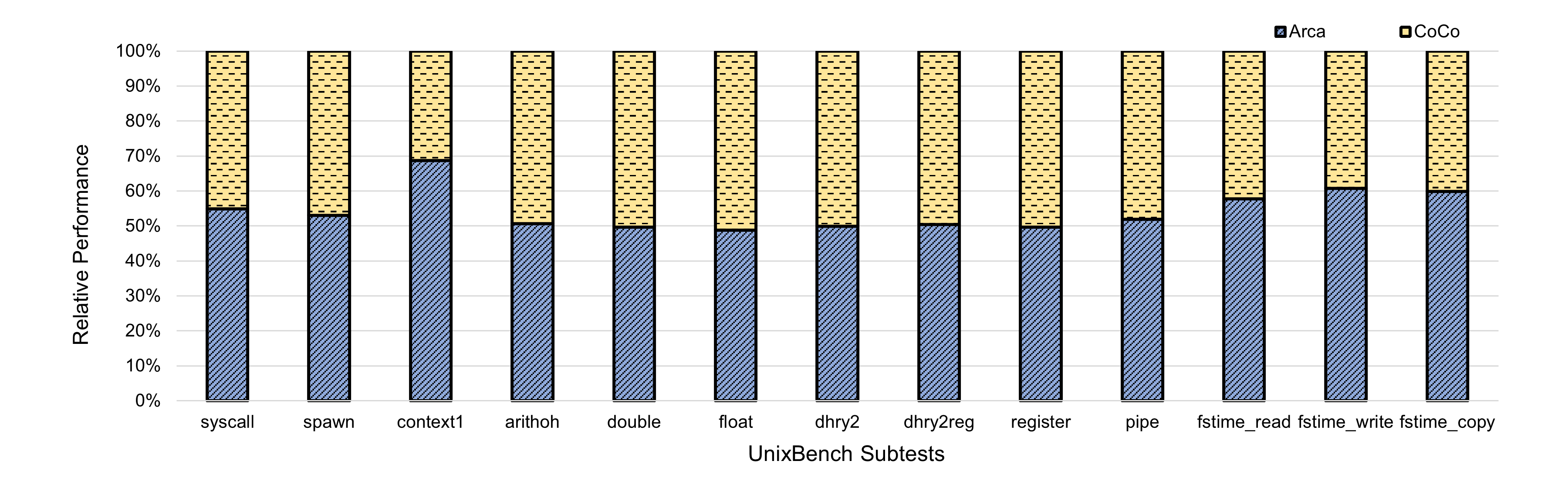}
    \caption{UnixBench performance comparison between \SYSNAME~and CoCo on AMD SEV. Both systems exhibit near-parity performance under SEV’s coarse-grained memory encryption, 
while \SYSNAME{} shows modest gains in \texttt{context1} and \texttt{fstime-*} subtests due to its direct kernel interaction in the TEE-in-Container model.}
    \label{fig:test_sev}
\end{figure*}
Figure~\ref{fig:test_sev} presents the UnixBench performance comparison between \SYSNAME~and CoCo on the AMD~SEV platform. Across all subtests, \SYSNAME delivers performance comparable to or slightly higher than CoCo, indicating that its TEE-in-Container architecture introduces negligible overhead under SEV's VM-granular memory encryption. The performance characteristics can be summarized into three major trends.

\textit{a) Execution efficiency: }
\SYSNAME~demonstrates a noticeable advantage in both the \texttt{context1} and \texttt{syscall} benchmarks. These tests are sensitive to kernel scheduling latency and syscall handling efficiency. In \SYSNAME, applications operate directly within the SEV-protected guest kernel, eliminating container-runtime mediation and namespace management that CoCo must still enforce inside its VM. This flattened execution structure reduces scheduling transitions and syscall validation overhead, resulting in faster context switching and more efficient kernel interaction. SEV’s hardware transparency further prevents additional trust-boundary operations, allowing the efficiency benefit of \SYSNAME{} to translate directly into measurable performance gains.

\textit{b) State synchronization efficiency: }
A consistent but modest performance advantage is observed in filesystem-related microbenchmarks such as \texttt{fstime\_read}, \texttt{fstime\_write}, and \texttt{fstime\_copy}. These workloads involve metadata updates and buffer synchronization within the trusted domain. Because system calls in \SYSNAME~are serviced directly by the SEV guest kernel, state transitions incur fewer mediation steps than in CoCo’s container-in-TEE design, where container-level policy checks remain active. Although SEV encrypts memory transparently at the memory controller, removing one layer of trusted mediation enables \SYSNAME~to achieve slightly higher throughput in I/O-triggered state transitions.

\textit{c) Neutralized execution paths: }
Other benchmarks—including \texttt{arithoh}, \texttt{float}, \texttt{double}, \texttt{dhry2}, \texttt{dhry2reg}, \texttt{register}, \texttt{pipe}, and \texttt{spawn}—exhibit near-identical results between \SYSNAME and CoCo. These workloads are dominated either by CPU/FPU pipelines or by in-guest communication primitives, both of which bypass inter-layer isolation control in SEV. Since SEV establishes trust at the VM level without necessitating enclave transitions or container-boundary revalidation, the architectural differences between the two designs become largely neutralized. Therefore, neither system incurs additional latency beyond the hardware-protected VM domain, resulting in performance parity across these execution categories.

\textit{Summary: }
\SYSNAME~achieves near-parity or slightly higher performance compared with CoCo across all UnixBench subtests on AMD SEV. The advantages in context switching and state-synchronization workloads validate that reducing software mediation within the trusted boundary improves efficiency without compromising SEV’s hardware-backed isolation guarantees. Meanwhile, computation- and communication-dominant workloads exhibit convergent performance due to SEV’s coarse-grained protection model. These results confirm that the TEE-in-Container architecture remains lightweight and practical when deployed on hypervisor-based TEEs such as SEV.

\section{Security Discussion}\label{sec:sec_dis}

Conventional confidential container systems, such as CoCo, encapsulate the entire container runtime and its supporting services inside a TEE-based VM. Although this approach prevents the host OS from directly accessing protected workloads, it inflates the TCB with a large amount of software that is irrelevant to the confidentiality and integrity of sensitive applications. The enlarged trusted code base introduces a broad attack surface and makes security verification and patch management significantly more challenging. In contrast, \SYSNAME~returns to the TEE's essential purpose by placing only the security-critical execution context in the hardware-protected domain and keeping container runtime components outside the TCB. As a result, \SYSNAME~achieves a more compact and auditable trusted environment, reduces the blast radius of potential compromises, and mitigates the risk posed by unnecessary privileged software within the TEE. In the rest of this section, we analyze how this architectural shift enhances confidentiality and integrity guarantees under the assumed threat model and achieves stronger security than the mainstream existing confidential container designs.

\textit{a) Architectural Security Implications}
\begin{figure}
    \centering
    \includegraphics[width=0.9\linewidth]{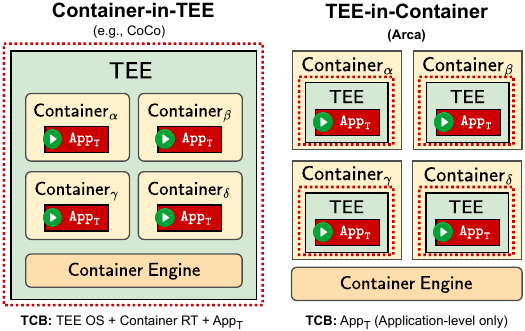}
    \caption{Comparison of Container-in-TEE (e.g., CoCo) and TEE-in-Container (\SYSNAME) architectures.
\SYSNAME~minimizes the trusted computing base by confining trust to application-level workloads.}
    \label{fig:arch-comp}
\end{figure}

The architectural contrast illustrated in Fig.~\ref{fig:arch-comp} highlights how \SYSNAME~fundamentally redefines the trust boundary of confidential container systems. Traditional \emph{Container-in-TEE} designs (e.g., CoCo, Kata Containers, and Enarx) encapsulate the entire container runtime and guest operating system within a TEE-based virtual machine. While this strategy successfully isolates workloads from an untrusted host, it inflates the TCB by incorporating non-essential components—such as container daemons, runtime shims, and kernel services—into the trusted domain. The enlarged TCB not only broadens the attack surface but also complicates attestation and formal verification, undermining the assurance that TEE protection intends to provide.

In contrast, \SYSNAME~adopts a \emph{TEE-in-Container} paradigm that decouples workload protection from system management. Each container instantiates its own hardware-isolated TEE dedicated solely to the confidential workload, while orchestration and runtime management remain outside the trusted boundary. This per-container TEE design eliminates shared runtime states, reduces inter-layer dependencies, and prevents privilege escalation across containers. As a result, the compromise of one container is strictly contained within its own TEE instance, preventing lateral movement or exposure of global secrets.

By shrinking the TCB to include only the application logic and its minimal security dependencies, \SYSNAME~restores the TEE’s essential purpose—providing a minimal, verifiable, and auditable protection boundary. This compact trust base not only improves the clarity of attestation evidence but also enhances the feasibility of formal verification and runtime integrity checking. Furthermore, it maintains compatibility with standard container orchestration workflows, ensuring that the gain in security does not come at the cost of operational flexibility. Overall, \SYSNAME~achieves a more balanced trade-off between confidentiality, manageability, and trust assurance in multi-tenant cloud-native environments.

\textit{b) Confidentiality Protection: Why plaintext never leaves the TEE? }

In \SYSNAME, all sensitive data and computations are confined within hardware-protected TEE domains, preventing any exposure to the untrusted host or container runtime. Memory encryption mechanisms provided by SGX, TDX, and SEV ensure that plaintext never leaves the enclave or the trusted VM boundary, even under host-level memory inspection (such as MCA attack), DMA probing, or snapshot attacks. Secrets are provisioned only after successful remote attestation, ensuring that decryption keys and sensitive data exist exclusively within verified and measured instances of \SYSNAME. Unlike CoCo, which embeds the container runtime and system services into the TEE and thus retains multiple code paths with potential data access privileges, \SYSNAME~limits trusted components to application-relevant code only. This minimalism substantially reduces the attack surface through which plaintext could be leaked or inferred.

\textit{c) Integrity Protection: Why can a host-level root not tamper with the execution state? }

\SYSNAME~binds each execution instance to cryptographic measurements performed at launch, ensuring that only verified code and configurations can run within the protected domain. Any attempt by the host OS or hypervisor to modify program binaries, memory pages, or control flow will trigger verification failures and prevent enclave or VM initialization. During runtime, nested page table protection (TDX/SEV) and enclave integrity checks (SGX) enforce strict separation between trusted and untrusted address spaces, rendering host-level write or remap operations ineffective. Moreover, outputs produced within the TEE can be sealed or signed with the attested identity, enabling downstream verifiers to detect any tampering or replay. Consequently, even a fully privileged host cannot alter or falsify the execution results of an application running in \SYSNAME.

\textit{d) Damage Containment vs. CoCo: Why per-container TEE minimizes the blast radius? }

Existing confidential container frameworks such as CoCo, Kata Containers, and Enarx typically deploy multiple containers within a single shared TEE instance. This coarse-grained protection model couples the fate of all co-located workloads---once the shared TEE is compromised through a runtime exploit or configuration error, every container within it becomes exposed. In contrast, \SYSNAME~follows a TEE-in-Container paradigm, where each container internally hosts its own TEE runtime dedicated to its specific confidential workload. This one-to-one binding between container and TEE creates mutually isolated trust domains that eliminate cross-tenant interference and privilege escalation. Even if an application within a single container is compromised, the attack surface is confined to that container’s hardware-protected boundary. Such fine-grained isolation not only constrains potential damage but also enhances accountability and auditability, aligning the trust granularity of \SYSNAME~with modern container orchestration semantics.

\textit{e) Host Compromise Resilience: Why is the threat model fully covered? }

The assumed adversary has complete control over the host operating system, including root privileges and direct access to the container manager. In this setting, conventional software-based isolation mechanisms become ineffective. \SYSNAME~eliminates dependency on host enforcement by removing the container runtime and kernel from the TCB. The host cannot observe or modify the execution context inside the TEE due to hardware-enforced memory protection and cryptographically bound attestation channels. All communication between the containerized application and external services occurs through authenticated, encrypted channels, ensuring that neither code nor data integrity is compromised in transit. Even under a compromised host, \SYSNAME~maintains the complete confidentiality and integrity of workloads, satisfying the protection requirements outlined in the threat model.

\textit{f) Comparative Security Landscape and Threat-to-Defense Mapping}

To provide a comprehensive view of how \SYSNAME~strengthens system security compared with existing confidential-container architectures, we extend our analysis from individual mechanisms to a cross-system comparison. Table~\ref{tab:comp_solutions} contrasts \SYSNAME~with representative designs such as CoCo, Kata Containers, and Enarx across multiple security dimensions, including TCB size, isolation granularity, attack surface, and operational manageability.

\begin{table*}[t]
\centering
\renewcommand{\arraystretch}{1.2}  
\setlength{\tabcolsep}{3pt}        
\caption{Comparison of \SYSNAME~with representative confidential-container systems.}
\label{tab:comp_solutions}
\begin{tabular*}{\textwidth}{@{\extracolsep{\fill}}%
>{\raggedright\arraybackslash}p{0.18\textwidth}  
>{\raggedright\arraybackslash}p{0.20\textwidth}  
>{\raggedright\arraybackslash}p{0.20\textwidth}  
>{\raggedright\arraybackslash}p{0.20\textwidth}  
>{\raggedright\arraybackslash}p{0.18\textwidth}@{}} 
\hline
\textbf{Dimension} & \textbf{\SYSNAME~(TEE-in-Container)} & \textbf{CoCo (Container-in-TEE)} & \textbf{Kata Containers} & \textbf{Enarx} \\
\hline
\textbf{TCB} &
Minimal; only security-critical application code resides in TEE. &
Large; includes full container runtime, OS services, and dependencies. &
Medium; includes lightweight guest OS and container agent. &
Medium; includes WebAssembly runtime and supporting libraries. \\

\textbf{Isolation Granularity} &
Per-container hardware isolation; each container embeds its own TEE instance. &
Shared TEE per VM; multiple containers share a single trusted domain. &
VM-level isolation per pod; partial container co-location possible. &
Process-level TEEs hosting multiple applications in one runtime. \\

\textbf{Cross-Tenant Impact Scope} &
Attack confined within one container; no shared TEE state or runtime. &
Compromise in one container may affect all in the same TEE. &
Limited by VM boundary but not per-container. &
Multiple workloads share the runtime—broader impact if compromised. \\

\textbf{Root-of-Trust Enforcement} &
Hardware-enforced TEE measurement; remote attestation gates secret provisioning. &
The TEE measurement includes a large software stack; verification is more complex. &
Relies on hypervisor + TEE VM chain; mixed trust anchors. &
Hardware attestation + WASM module verification. \\

\textbf{Attack Surface Inside TEE} &
Minimal—no init system, package manager, or container daemon. &
Large—includes containerd, systemd, and library dependencies. &
Moderate—depends on guest OS and container manager. &
Moderate—WASM runtime complexity inside TEE. \\

\textbf{Confidentiality Guarantee} &
Strong; plaintext never leaves the TEE-in-Container domain. &
Strong; but weakened by larger TCB exposure. &
Strong; but sensitive to guest-OS vulnerabilities. &
Strong; data confined to WASM enclave, but limited I/O model. \\

\textbf{Integrity Guarantee} &
Measured launch + hardware page protection; per-container verification. &
Measured launch but shared runtime allows cross-impact. &
Integrity depends on guest kernel + TEE co-validation. &
Relies on WASM runtime integrity enforcement. \\

\textbf{Operational Manageability} &
Managed as regular containers; compatible with orchestration (e.g., Kubernetes). &
Requires modified container orchestration to manage in-TEE runtime. &
Natively supported via CRI plugin integration. &
Requires WASM toolchain and a specific build pipeline. \\

\textbf{Typical Use Cases} &
Fine-grained confidential workloads, cloud-edge cooperation, and multi-tenant isolation. &
General-purpose container protection in IaaS environments. &
Lightweight confidential VMs for cloud-native workloads. &
WASM-based secure application execution. \\
\hline
\end{tabular*}
\end{table*}

From the comparison, several key observations can be made. 
First, \SYSNAME~achieves the smallest TCB among all systems by adopting a \emph{TEE-in-Container} paradigm that encapsulates only security-critical components within a hardware-protected domain, avoiding the inclusion of container runtimes, guest OS services, or orchestration daemons. 
Second, this fine-grained per-container protection model eliminates the shared-trust dependency inherent in \emph{Container-in-TEE} solutions (e.g., CoCo), thereby preventing lateral movement or cross-tenant privilege escalation once a single container is compromised. 
Third, the attack surface inside each TEE is drastically reduced—no \texttt{systemd}, package managers, or container daemons exist within the trusted boundary—simplifying both attestation and runtime verification. 
Finally, because each container directly manages its TEE instance via standard container interfaces, \SYSNAME~preserves the deployment and orchestration compatibility of conventional container ecosystems such as Kubernetes, combining hardware-rooted trust with cloud-native manageability.

\begin{table}[t]
\centering
\renewcommand{\arraystretch}{1.15}
\setlength{\tabcolsep}{5pt}
\caption{Mapping between identified threats and \SYSNAME~defenses.}
\label{tab:threat_defense}
\begin{tabular}{p{0.28\linewidth} p{0.64\linewidth}}
\toprule
\textbf{Threat Type} & \textbf{\SYSNAME~Defense and Mechanism} \\
\midrule
Host-level memory disclosure &
Memory encryption (SGX EPC, TDX EPT, SEV encrypted memory) prevents plaintext leakage through memory dumps, DMA, or snapshots. Secrets are provisioned only after successful attestation. \\

Code or data tampering by the host &
Measured launch ties execution to verified binaries. Unauthorized modification of memory pages or control flow triggers a hardware validation failure. Outputs can be sealed or signed. \\

Cross-container lateral attack &
Each container hosts its own TEE instance, forming isolated trust domains. Compromise in one container cannot propagate to others. \\

Untrusted OS/VMM interference &
The host and hypervisor are excluded from the TCB. Hardware isolation prevents privileged manipulation of TEE state or registers. \\

Misconfiguration inside TEE &
\SYSNAME's minimal TCB excludes auxiliary services and daemons, reducing exploitable configuration complexity. \\
\bottomrule
\end{tabular}
\vspace{-0.5em}
\end{table}


To connect this architectural comparison to the defined threat model, Table~\ref{tab:threat_defense} summarizes how \SYSNAME~neutralizes each attack class. 
All threats to confidentiality and integrity are mitigated through hardware isolation, attestation-gated provisioning, and per-container TEE segmentation, confirming that the system provides end-to-end resilience under the assumed adversarial conditions.

\textit{g) Verifiability and Scalability}

Finally, the modular design of \SYSNAME~enhances the verifiability and scalability of trusted computing in cloud-native environments. 
Because each container hosts a self-contained and independently attestable TEE, formal verification and evidence validation can be performed per-container without requiring global synchronization. 
This property makes \SYSNAME~particularly suitable for large-scale multi-tenant deployments where both strong isolation assurance and efficient trust auditing are critical.

\section{Related Work}\label{sec:rel_work}


Containers have become the cornerstone of cloud-native computing, enabling lightweight virtualization and efficient deployment of applications at scale. However, traditional containers inherently rely on the host operating system for isolation, which exposes them to various security threats such as kernel-level attacks, malicious administrators, and cross-tenant data leakage in multi-tenant environments. To mitigate these security risks, researchers have proposed various approaches to enhance container isolation and data protection, which can be broadly categorized into two classes: software-based protection and TEE-based confidential computing schemes.

Software-based approaches have been proposed to enhance container isolation without requiring hardware-based security support or modifications, among which gVisor\cite{gVisor}-\cite{gVisorbad}, Kata Containers\cite{Kata}-\cite{Katatest}, and Nabla Containers\cite{Nabla} are representative examples. gVisor implements a user-space kernel that intercepts and handles application system calls in a sandboxed environment. By minimizing direct interactions with the host kernel, it effectively reduces the attack surface while maintaining good compatibility with existing container ecosystems. However, its user-space system call emulation may introduce non-negligible performance overhead. Kata Containers represents an intermediate approach between traditional containers and full virtual machines. It launches each container inside a lightweight VM, combining hardware-assisted virtualization with the container runtime interface. This design significantly improves isolation and security compared to standard containers, while preserving the deployment flexibility of container orchestration systems. Nonetheless, it still depends on the hypervisor’s trustworthiness and cannot fully protect against privileged host attacks. Nabla Containers take a different approach by leveraging unikernel-like techniques to minimize the interface between the application and the host. Built upon a library OS, Nabla Containers provide a small, well-defined system call boundary, thus reducing the potential attack surface. Yet, due to its specialized runtime model, Nabla suffers from limited compatibility with existing containerized applications.

Although these software-based mechanisms offer improved isolation and deployment flexibility, they still fundamentally rely on the host kernel or hypervisor for resource management. As a result, their security guarantees are insufficient in untrusted or multi-tenant cloud environments where privileged adversaries may compromise the host. 

To address the inherent trust dependency of software-based protection approaches, TEE-based confidential containers (as the mainstream hardware-supported security mechanism) have emerged as a promising solution for secure container execution in untrusted clouds. These approaches leverage hardware-assisted TEEs such as Intel SGX, AMD SEV, and Intel TDX to provide runtime isolation and hardware-level memory encryption, ensuring that sensitive code and data remain protected even from privileged attackers.

Brasser et al. \cite{Brasser2022} proposed the Trusted Container Extensions (TCX) architecture, which combines the flexible management of standard containers with the hardware-based protection provided by AMD SEV. Johnson et al.\cite{Johnson2023}-\cite{Johnson2024} introduced the Parma architecture that leverages SEV-SNP to enable hardware-enforced confidential computing for containers. Hua et al. \cite{Hua2021} designed the TZ-Container mechanism, which uses ARM TrustZone to create multiple isolated execution environments to defend against emerging container attacks. Van’t Hof et al. \cite{Van‘tHof2022} proposed the BlackBox architecture, which employs ARM hardware virtualization and nested paging to achieve fine-grained data protection. Arnautov’s SCONE\cite{Arnautov2016} and Lee’s TeeMate\cite{Lee2024} fully exploit Intel SGX capabilities to enhance container security while improving performance and reducing the TCB size. Compared with the above studies, existing approaches have made significant progress in improving container security but still exhibit certain limitations. Most of these solutions rely on specific hardware platforms and lack cross-platform compatibility, limiting their deployment and adoption across diverse cloud computing environments.

The open-source project CoCo\cite{CoCo} proposes a unified framework for cloud-native confidential computing by integrating TEE capabilities into mainstream container runtimes such as \texttt{containerd} and \texttt{CRI-O}. This design protects sensitive data and code within containerized workloads and enables rapid deployment across multiple hardware platforms. Although CoCo provides broader hardware compatibility, running entire containers inside the TEE significantly enlarges the TCB, introducing additional security and performance challenges. The expansion of the TCB not only increases the system’s attack surface and reduces overall security but also incurs additional performance overhead, which conflicts with the cloud-native requirements for high performance, low latency, and lightweight deployment, and deviates from the original intent of TEE design—to confine only security-critical computations within a minimal, verifiable trusted domain rather than encapsulating the whole container runtime and its ancillary libraries.

Overall, these studies have made significant progress in enhancing the security of container runtimes. However, challenges remain in cross-platform compatibility, performance optimization, and TCB control. Achieving secure isolation while minimizing the TCB size, reducing performance overhead, and improving the usability of confidential containers across diverse computing environments remains a key open problem in cloud-native confidential computing.

\section{Conclusion}\label{sec:concl}
This paper presented \SYSNAME, a novel confidential-container architecture that redefines the trust boundary between containers and trusted execution environments. 
Unlike conventional ``Container-in-TEE'' approaches such as CoCo, which encapsulate the entire container runtime within the TEE and suffer from inflated TCBs and enlarged attack surfaces, \SYSNAME~follows a ``TEE-in-Container'' paradigm. 
In this design, each container hosts an independently instantiated and attested TEE instance that protects only the security-critical execution context, while keeping container orchestration and runtime services outside the trusted domain. 
This architectural inversion significantly reduces the TCB size, enhances auditability, and aligns the granularity of trust isolation with that of container management.

We implemented \SYSNAME~on three representative hardware TEE platforms—Intel SGX, Intel TDX, and AMD SEV—and conducted comprehensive evaluations using both micro-benchmarks and system-level workloads. 
The experimental results demonstrate that \SYSNAME~achieves comparable or superior performance to CoCo while maintaining strong confidentiality and integrity guarantees. 
Specifically, under the TDX and SEV platforms, \SYSNAME~shows consistent advantages in syscall, arithmetic, and context-switch benchmarks due to its simplified execution hierarchy and minimized boundary crossings. 
Security analysis further confirms that \SYSNAME~retains full protection coverage against host-level privilege compromise and cross-container attacks, achieving end-to-end resilience with a smaller TCB.

In summary, \SYSNAME~bridges the gap between cloud-native container management and hardware-assisted trusted computing by offering a unified, efficient, and scalable way to deploy per-container TEEs without modifying existing orchestration frameworks. 
Future work will extend \SYSNAME{} toward cross-TEE interoperability and formal verification, aiming to build a portable and verifiable foundation for confidential computing across heterogeneous trusted hardware ecosystems.

\section*{Acknowledgment}
The authors would like to thank the editor-in-chief, associate editor, and reviewers for their valuable comments and suggestions. This research was supported by the National Natural Science Foundation of China (62572377, 62220106004). the 
This paper is supported by the National Key R\&D Program of China (No. 2023YFB3107500), the Innovation Capability Support Program of Shaanxi (No. 2023-CX-TD-02), the Xidian University Specially Funded Project for Interdisciplinary Exploration (No. TZJHF202502) and the Fundamental Research Funds for the Central Universities (No. ZDRC2202).

\begin{IEEEbiography}[{\includegraphics[width=1in,height=1.25in,clip,keepaspectratio]{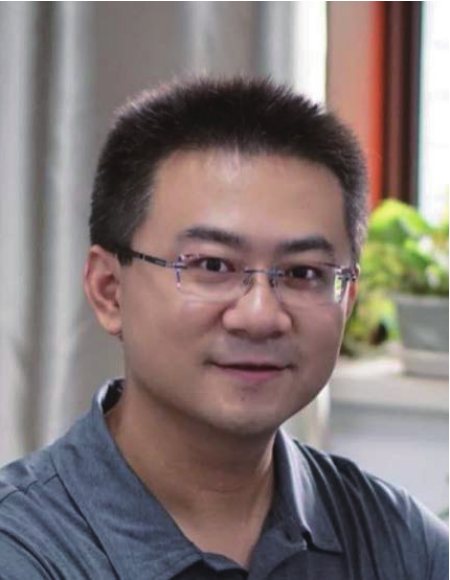}}]{Di Lu}
(Member, IEEE) received the B.S., M.S., and Ph.D. degrees in computer science and technology from Xidian University, China, in 2006, 2009, and 2014. Now he is a  Professor in the School of Computer Science and Technology at Xidian University. His research interests include trusted computing, confidential computing, system and network security.
\end{IEEEbiography}

\begin{IEEEbiography}[{\includegraphics[width=1in,height=1.25in,clip,keepaspectratio]{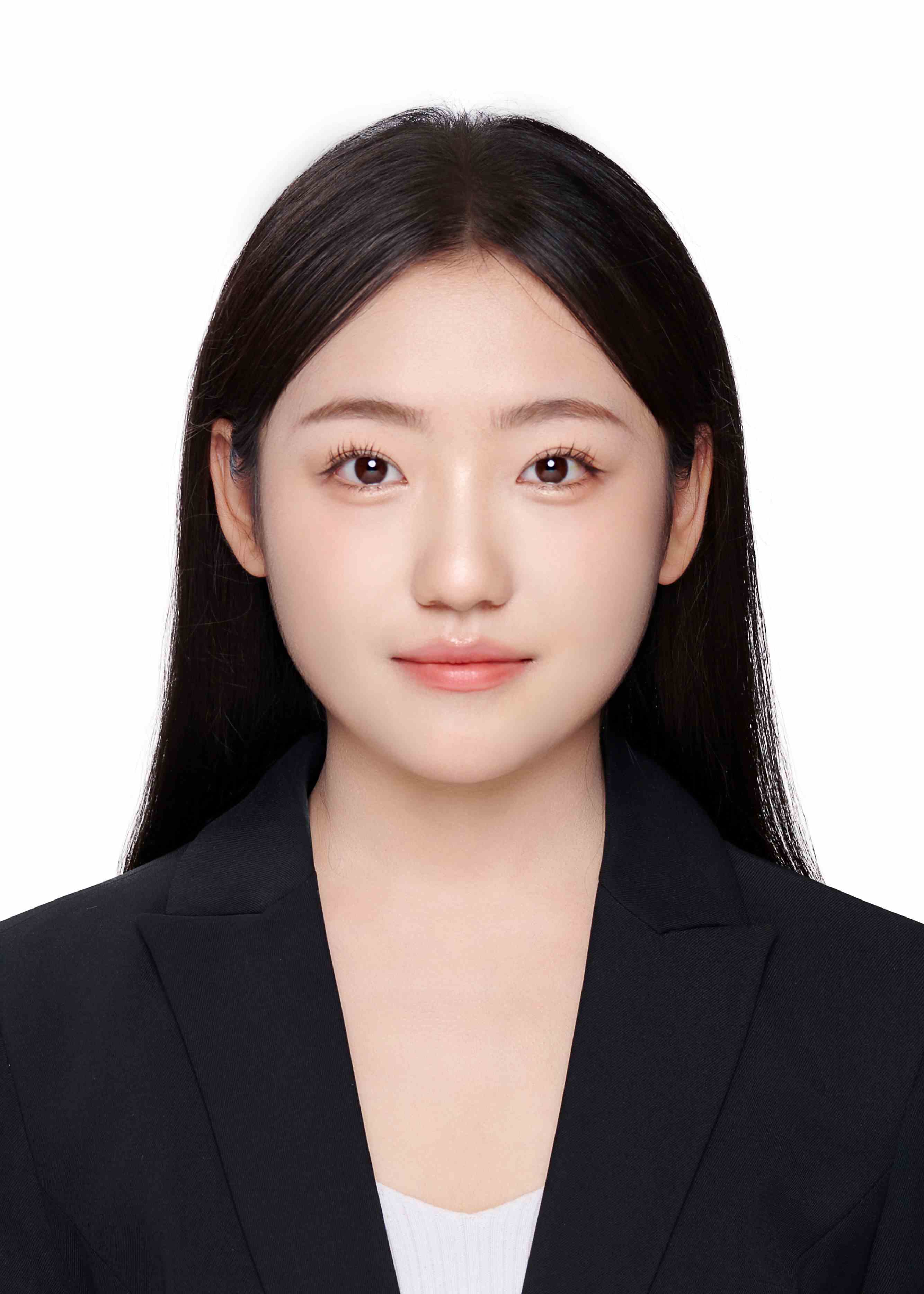}}]{Mengna Sun}
received the BS degree from Hebei University of Technology, China, in 2022. She is currently pursuing an MS degree in the School of Computer Science and Technology at Xidian University, China. Her research interests include trusted computing and embedded system security.
\end{IEEEbiography}

\begin{IEEEbiography}[{\includegraphics[width=1in,height=1.25in,clip,keepaspectratio]{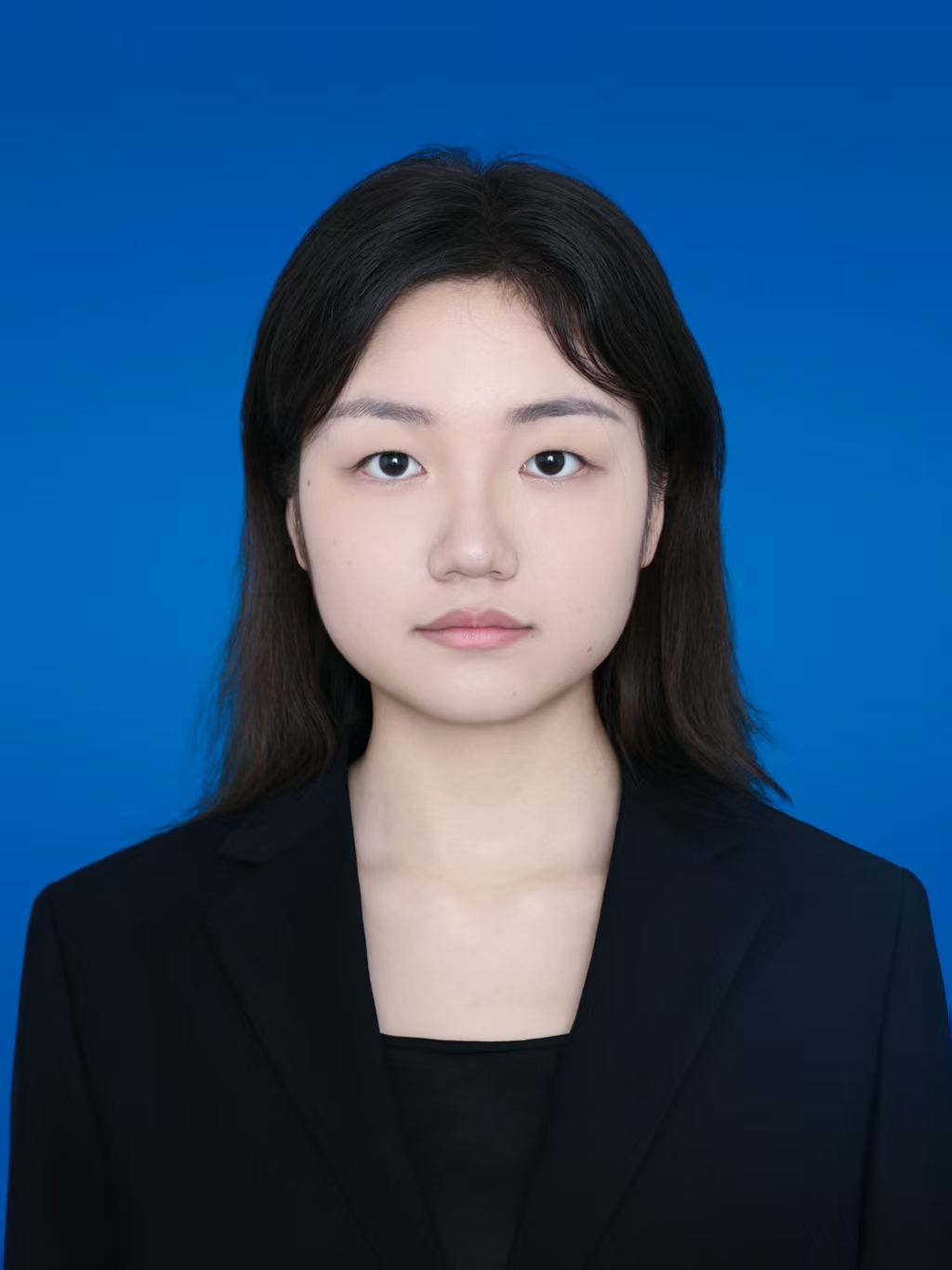}}]{Qingwen Zhang}
received the BS degree from Xidian University, China, in 2024. She is currently pursuing an MS degree in the School of Computer Science and Technology at Xidian University, China. Her research interests include trusted computing and embedded system security.
\end{IEEEbiography}

\begin{IEEEbiography}[{\includegraphics[width=1in,height=1.25in,clip,keepaspectratio]{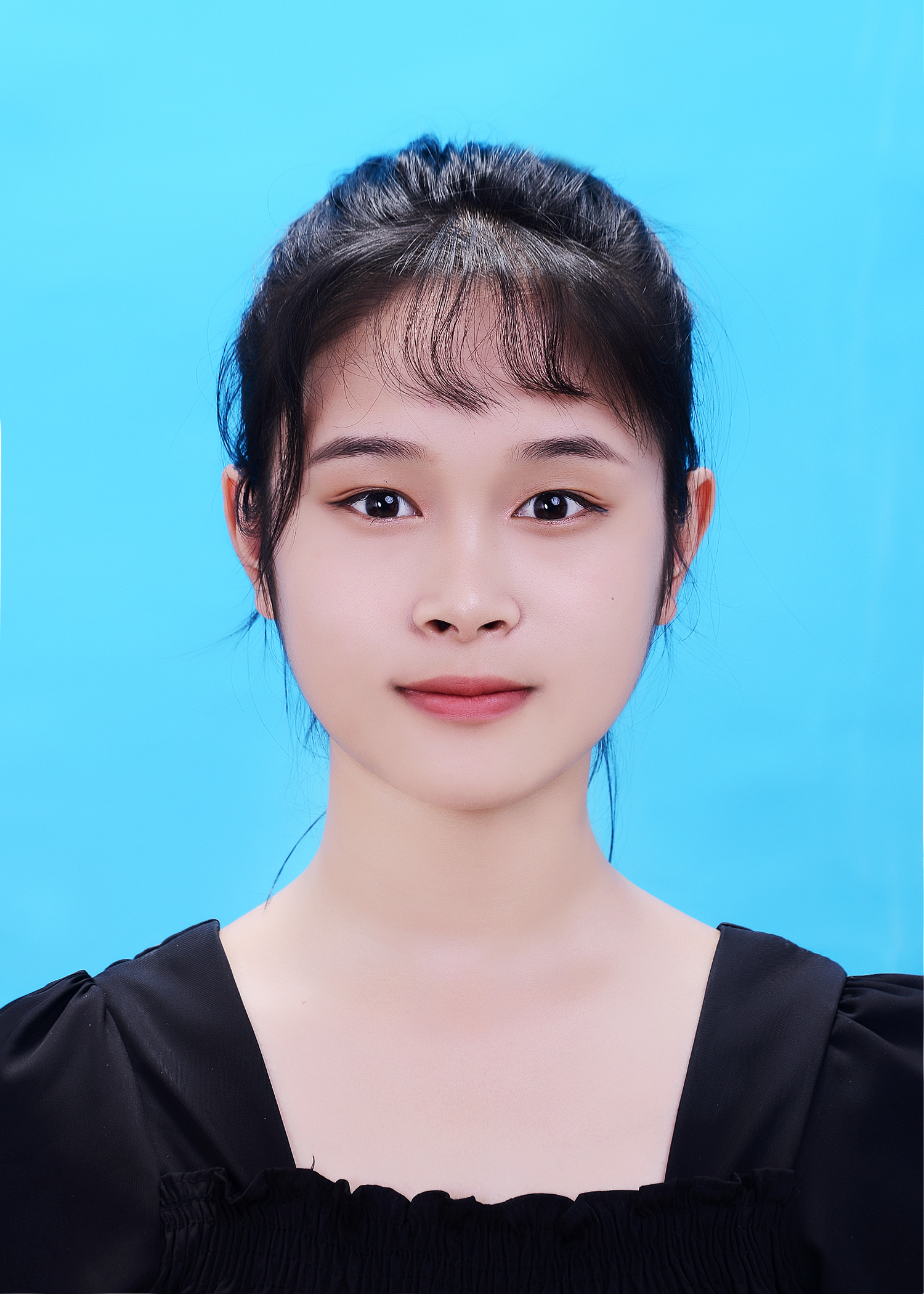}}]{Yujia Liu}
received the BS degree from Xi’an Shiyou University, China, in 2025. She is currently pursuing an MS degree in the School of Computer Science and Technology at Xidian University, China. Her research interests include trusted computing and embedded system security.
\end{IEEEbiography}

\begin{IEEEbiography}[{\includegraphics[width=1in,height=1.25in,clip,keepaspectratio]{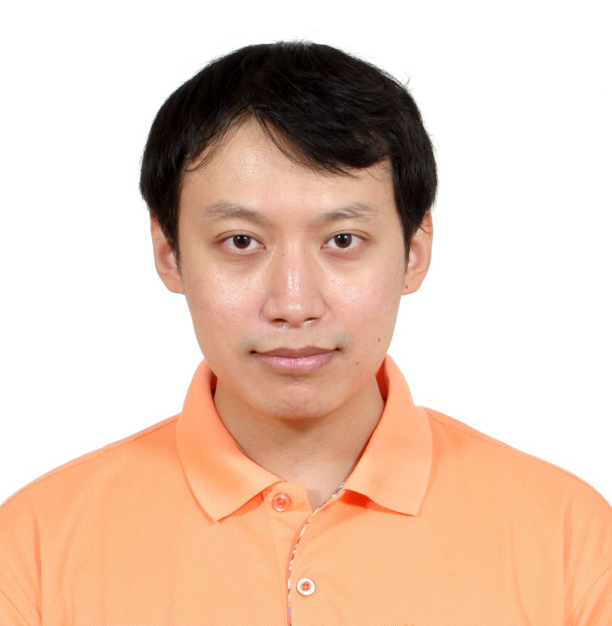}}]{Jia Zhang}
is a Confidential Computing Architect at Alibaba Cloud, specializing in the technical field of integrating confidential computing with container runtimes and AI data security. Joined Alibaba in 2017 and currently focuses on operating system security within Alibaba Cloud, including areas such as trusted computing and confidential computing.
\end{IEEEbiography}

\begin{IEEEbiography}[{\includegraphics[width=1in,height=1.25in,clip,keepaspectratio]{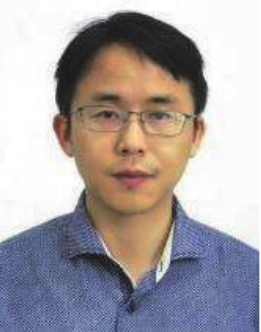}}]{Xuewen Dong}
(Member, IEEE) received the B.E., M.S., and Ph.D. degrees in computer science and technology from Xidian University, Xi’an, China, in 2003, 2006, and 2011, respectively. From 2016 to 2017, he was a Visiting Scholar with Oklahoma State University, Stillwater, OK, USA. Currently, he is a Professor with the School of Computer Science and Technology, Xidian University. His research interests include blockchain and the security of smart systems.
\end{IEEEbiography}

\begin{IEEEbiography}[{\includegraphics[width=1in,height=1.25in,clip,keepaspectratio]{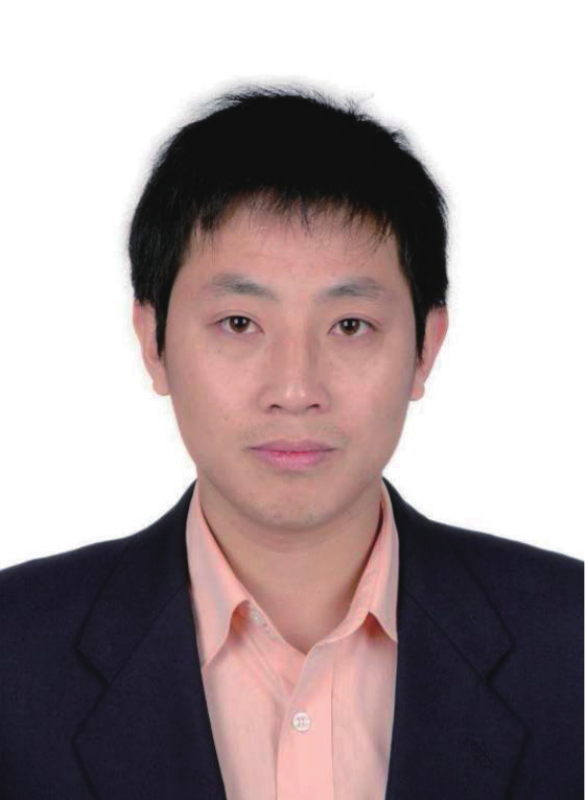}}]{YuLong Shen}
(Member, IEEE) received the BS and MS degrees in computer science and a PhD degree in cryptography from Xidian University, Xi’an, China, in 2002, 2005, and 2008, respectively. He is currently a professor with the School of Computer Science and Technology, Xidian University. His research interests
include wireless network security and cloud computing security.
\end{IEEEbiography}

\begin{IEEEbiography}[{\includegraphics[width=1in,height=1.25in,clip,keepaspectratio]{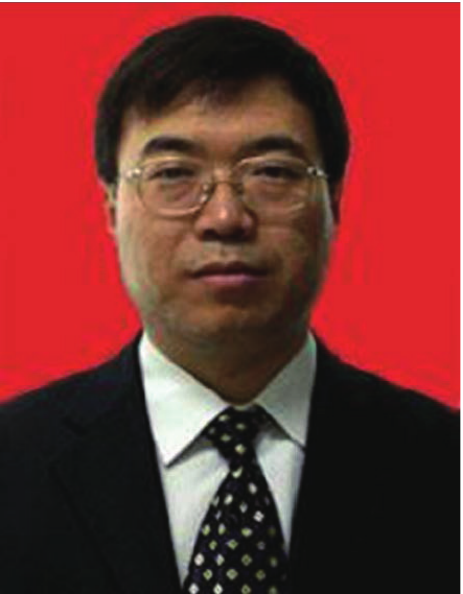}}]{Jianfeng Ma}
(Member, IEEE) received the BS degree in mathematics from Shaanxi Normal University, China, in 1985, and the MS and PhD degrees in computer software and communications engineering from Xidian University, China, in 1988 and 1995, respectively. Now, he is a professor with the School of Cyber Engineering, Xidian University, China. His current research interests include distributed systems, computer networks, and information and network security.
\end{IEEEbiography}

\vspace{11pt}

\vfill

\end{document}